\documentclass[iop,floatfix,numberedappendix,twocolappendix]{emulateapj} 

\usepackage[backref,breaklinks,colorlinks,urlcolor=blue,citecolor=blue,linkcolor=blue]{hyperref}
\usepackage[all]{hypcap}

\usepackage{graphicx}
\usepackage{enumerate}
\usepackage{amsmath,amssymb}
\usepackage{bm}
\usepackage{color}
\usepackage[utf8]{inputenc}

\newcommand{\vM}{\mathsf{M}}
\newcommand{\vD}{\mathsf{D}}
\newcommand{\vR}{\mathsf{R}}
\newcommand{\vC}{\mathsf{C}}

\newcommand{\trans}{\mathsf{T}}
\newcommand{\teff}{T_\textrm{eff}}
\newcommand{\logg}{\log g}
\newcommand{\Z}{[{\rm Fe}/{\rm H}]}

\newcommand{\vsini}{v \sin i}
\newcommand{\matern}{Mat\'{e}rn}

\newcommand{\flam}{f_\lambda}
\newcommand{\vt}{ {\bm \theta}}
\newcommand{\vT}{ {\bm \Theta}}
\newcommand{\vp}{ {\bm \phi}}
\newcommand{\vP}{ {\bm \Phi}}
\newcommand{\cheb}{ \vp_{\mathsf{P}}}

\newcommand{\cov}{ \vp_{\mathsf{C}}}

\newcommand{\KK}{\mathcal{K}}
\newcommand{\Kglobal}{\KK^{\textrm{G}}}
\newcommand{\Klocal}{\KK^{\textrm{L}}}

\newcommand{\PHOENIX}{{\sc Phoenix}}

\newcommand{\wg}{\mathbf{w}^\textrm{grid}}
\newcommand{\wgh}{\hat{\mathbf{w}}^\textrm{grid}}

\newcommand{\Sg}{\mathbf{\Sigma}^\textrm{grid}}

\shorttitle{Flexible Spectroscopic Inference}
\shortauthors{Czekala et al.}

\begin{document}

\title{Constructing a Flexible Likelihood Function for Spectroscopic Inference}

\newcommand{\harvard}{1}
\newcommand{\nyu}{2}
\newcommand{\mpia}{3}
\newcommand{\cds}{4}

\author{%
  Ian~Czekala\altaffilmark{\harvard }, 
  Sean~M.~Andrews\altaffilmark{\harvard },
  Kaisey~S.~Mandel\altaffilmark{\harvard },
  David~W.~Hogg\altaffilmark{\nyu },
  and Gregory~M.~Green\altaffilmark{\harvard}
}

\altaffiltext{\harvard}  {Harvard-Smithsonian Center for Astrophysics, 
			 60 Garden Street, Cambridge, MA 02138; \email{iczekala@cfa.harvard.edu}}
\altaffiltext{\nyu}      {Center for Cosmology and Particle Physics,
                          Department of Physics, New York University,
                          4 Washington Place, New York, NY, 10003, USA}

\begin{abstract}
We present a modular, extensible likelihood framework for spectroscopic inference  based on synthetic model spectra.  The subtraction of an imperfect model from a continuously sampled spectrum introduces covariance between adjacent datapoints (pixels) into the residual spectrum.  For the high signal-to-noise data with large spectral range that is commonly employed in stellar astrophysics, that covariant structure can lead to dramatically underestimated parameter uncertainties (and, in some cases, biases).  We construct a likelihood function that accounts for the structure of the covariance matrix, utilizing the machinery of Gaussian process kernels.  This framework specifically address the common problem of mismatches in model spectral line strengths (with respect to data) due to intrinsic model imperfections (e.g., in the atomic/molecular databases or opacity prescriptions) by developing a novel local covariance kernel formalism that identifies and self-consistently downweights pathological spectral line ``outliers."  By fitting many spectra in a hierarchical manner, these local kernels provide a mechanism to learn about and build data-driven corrections to synthetic spectral libraries.  An open-source software implementation of this approach is available at \url{http://iancze.github.io/Starfish}, including a sophisticated probabilistic scheme for spectral interpolation when using model libraries that are sparsely sampled in the stellar parameters.  We demonstrate some salient features of the framework by fitting the high resolution $V$-band spectrum of WASP-14, an F5 dwarf with a transiting exoplanet, and the moderate resolution $K$-band spectrum of Gliese~51, an M5 field dwarf. 
\end{abstract}

\section{Introduction} \label{sec:intro}

All astronomers recognize that spectroscopy offers a wealth of information that can help characterize the properties of the observing target.  In the context of stellar astrophysics, spectroscopy plays many fundamental roles.  The relative strengths and widths of stellar absorption lines provide access to physical properties like effective temperature ($\teff$) and surface gravity ($\logg$), enabling model comparisons in the Hertzsprung-Russell diagram to estimate the masses and ages so crucial to understanding stellar evolution, as well as individual elemental abundances or the collective ``metallicity" (typically parameterized as $\Z$), facilitating studies of the chemical hallmarks of different stellar populations.  With sufficient resolution, a spectrum also conveys information about rotation ($ \vsini$) and kinematics (e.g., association with a cluster or companion through the radial velocity, $v_r$).  While many fields benefit from such spectroscopic measurements, they are of acute interest to the exoplanet community.  There, all estimates of the planet properties are made {\it relative} to the host properties (e.g., the mass function and planet-to-host radius {\it ratio} are constrained with the radial velocity or transit techniques, respectively).  Moreover, essential clues to the planet formation process are encapsulated in the dependences of planet frequency on host mass \citep[e.g.,][]{johnson07,howard10} and metallicity \citep[e.g.,][]{fischer05,buchhave14}.   

The robust and quantitative extraction of physical (or empirical) parameters from an observed spectrum can be an extraordinary challenge.  Stellar models serve as comparative benchmarks to associate observed spectral features with the parameters of interest.  Generating a synthetic model spectrum involves a complex numerical treatment of the stellar structure and radiative transfer through the atmosphere \citep[e.g.,][]{kurucz93,castelli04, hauschildt99,husser13,paxton11}.  Detailed models calibrated to individual stars are important, but rare (e.g., the Sun, Vega); therefore, these stellar models are relatively untested in large swaths of parameter-space.  Moreover, they necessarily include simplifications to treat complicated physical processes (e.g., convection) or computational limitations (e.g., boundary conditions), and often must rely on incomplete or inaccurate atomic and molecular information (e.g., opacities).  In principle, the models could be improved with appropriate reference to spectroscopic data.  Nevertheless, they are remarkably successful in reproducing many diagnostic spectral features.  

There are various well-tested approaches being used in stellar astrophysics to compare these models with observed spectra and thereby infer basic parameters.  Perhaps the most common is a straightforward empirical technique that relies on distilling an information-rich subset of the data, usually in the form of spectral line equivalent widths and/or local continuum shapes.  A combined sequence of the ratios of these quantities can be especially sensitive to a given model parameter \citep[e.g., {\tt MOOG};][]{sneden73,gray94,reid95,rojas-ayala10,rojas-ayala12}.  This ``indexing" approach has the advantage of being trivially fast.  But, each condensed relationship is only informative over a limited swath of parameter-space, and it potentially masks degeneracies that are encoded in the spectral line shapes.  Another standard approach exploits the cross-correlation of an observed spectrum with a suite of model templates to optimize a set of parameters, usually with some weighting applied to specific spectral regions \citep[e.g., {\tt SPC};][]{buchhave12}.  In this case, the speed advantage is maintained (perhaps enhanced) and more data content is used (particularly in the spectral dimension), thereby achieving higher precision even for data with comparatively low signal-to-noise.  The disadvantage is that the model quality and parameter inferences are assessed in a heuristic (rather than probabilistic) sense, making it difficult to quantify uncertainty in the stellar parameters. A more direct method employs a pixel-by-pixel comparison between model and data. This has the benefits of increased parametric flexibility (e.g., one can fit for arbitrary abundances or structures) and a proper inference framework \citep[usually a least-squares approach, although increasingly in a Bayesian format;][]{shkedy07, schonrich14}. Ultimately, rather than pre-computing a library of sythetic spectra, one would like to incorporate the spectral synthesis back-end (e.g., {\tt SME}; \citealt{valenti96}) directly into the likelihood function, bypassing any interpolation when assessing the fit of stellar parameters in-between grid points in the library. Unfortunately, this is not yet computationally feasible beyond a limited wavelength range.

In this article, we construct a flexible forward-modeling method for the general spectroscopic inference problem in a Bayesian framework, building on the best aspects of the latter two approaches highlighted above. The key developments in this design include a spectral emulator to address the difficult task of interpolation in coarsely sampled synthetic spectral libraries and a non-trivial covariance matrix parameterized by both global (stationary) and local (non-stationary) Gaussian process kernels.  When combined with an appropriately sophisticated set of quantitative metrics for the relevant physical parameters, this method will efficiently propagate systematic uncertainties into the parameter inferences.  Ultimately, this approach could be employed to leverage spectroscopic data as a reference for improving the models.  

A complete overview of the methodology behind this approach is provided in Section \ref{sec:method}.  Some tests and example applications (for a high resolution optical spectrum of an F star, and a medium-resolution near-infrared spectrum of a mid-M star) are described in Section \ref{sec:examples}.  Finally, a discussion of its potential utility, especially the possibility of extending it to develop data-driven spectral models, is provided in Section \ref{sec:discussion}.  \\

\section{Methodology} \label{sec:method}

This section describes a generative Bayesian modeling framework that confronts some of the key technical obstacles in the spectroscopic inference problem.  The goal is to conservatively extract the maximal amount of information about a prescribed (and usually degenerate) parameter set by forward-modeling an observed spectrum, while also recognizing and explicitly accounting for the covariances (and potentially biases) introduced by pathologically imperfect models.  The method is modular, and therefore can easily incorporate additional physical or nuisance parameters as desired without sacrificing an accurate reflection of the limitations in the data.  The specific applications discussed here are related to the spectra of individual stars, but the methodology is generic (and could be used for the composite spectra of unresolved stellar clusters, galaxies, etc.).  

\begin{figure}[!b]
  \includegraphics{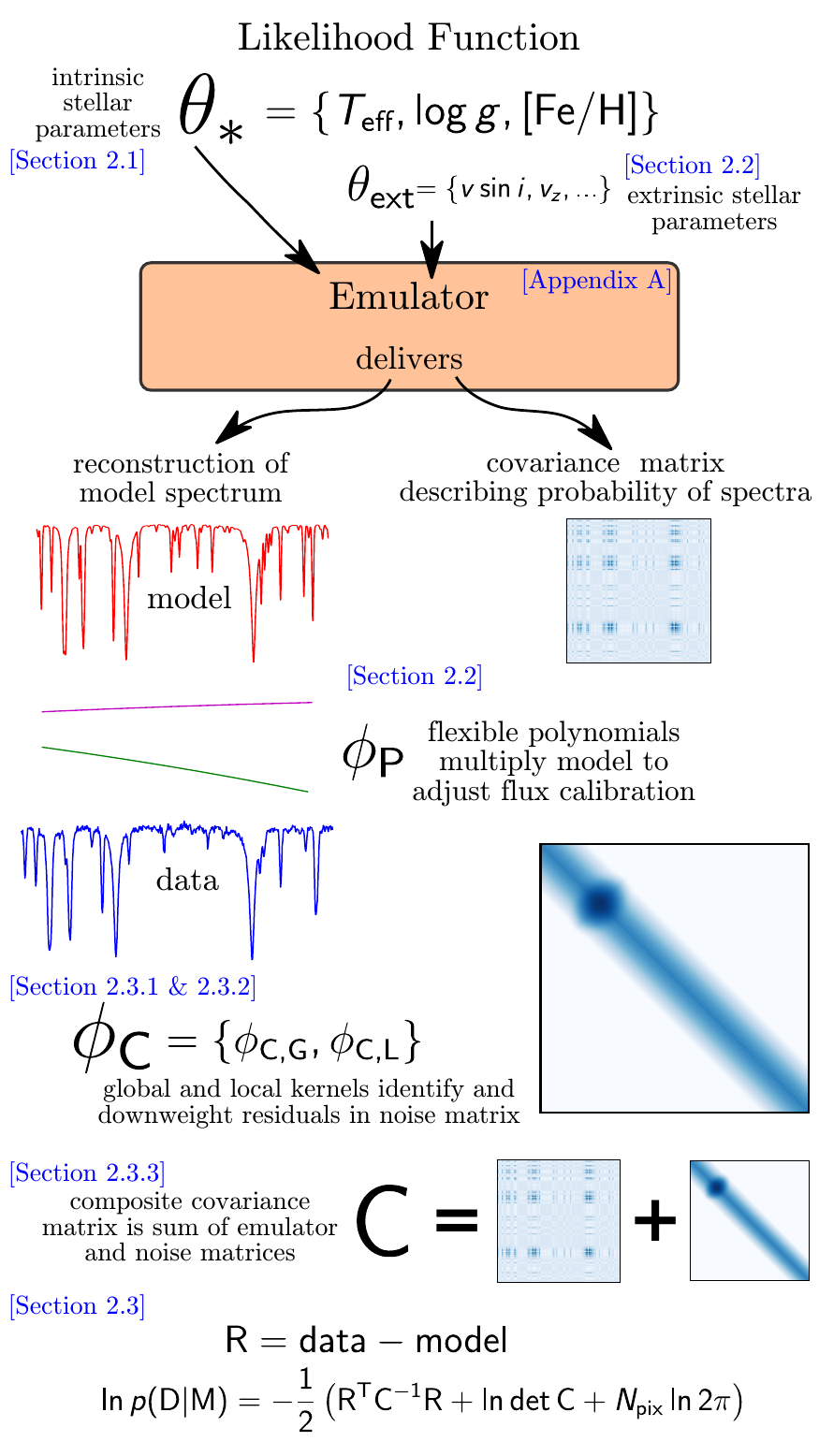}
  \figcaption{A flowchart showing how the parameters of the model are combined to forward model a spectrum. Before starting inference for a particular star, a Bayesian emulator is tuned to efficiently interpolate a grid of synthetic spectra (Appendix \ref{sec:Appendix}) for any queried set of ``intrinsic" stellar parameters ($\vt_{\ast}$). The spectrum is then modified according to ``extrinsic'' stellar parameters ($\theta_\textrm{ext}$) like $\vsini$ and $v_r$. Then, calibration polynomials ($\phi_\mathsf{P}$) provide slight adjustment to the continuum shape of the model to account for uncertainties in flux calibration. The second major component of the framework is accounting for covariant residual structure by using kernels to set the structure of the ``noise" matrix to downweight erroneous residual structure. Then, the multidimensional likelihood function is evaluated using the sum of these covariance matrices.
\label{fig:flowchart}}
\end{figure}

Figure~\ref{fig:flowchart} serves as a graphical guide to the mechanics of this modeling framework, and the remainder of this section.  First, a model spectrum is generated for a given set of physical parameters (Section~\ref{subsec:synthetic}; Appendix~\ref{sec:Appendix}), and then post-processed to mimic reality using a set of observational and practical nuisance parameters (Section~\ref{subsec:postprocess}).  Next, a direct, pixel-by-pixel comparison between the data and model spectra is made with a prescribed likelihood function and a parametric treatment of the covariances between pixel residuals (Section~\ref{subsec:likelihood}).  That process is iterated using Markov Chain Monte Carlo (MCMC) simulations in a multi-stage Gibbs sampler to numerically explore the posterior probability density of the model conditioned on the data, and thereby to determine constraints on the parameters of interest (Section~\ref{subsec:MCMC}).  Along the way, these procedures are illustrated with observations of the high resolution optical spectrum from a nearby F star.  That specific application, along with some alternative demonstrations of the method, are discussed in more detail in Section~\ref{sec:examples}.

\subsection{Generating a Model Spectrum \label{subsec:synthetic}}

There are many approaches for generating a model spectrum, $f_{\lambda}$, for a specific set of parameters, $\vt_{\ast} = \{\teff, \logg, \Z \}$.  In the most direct case of spectral synthesis, a model atmosphere structure is assembled and simulations of energy transport through it are conducted with a radiative transfer code \citep[e.g.,][]{kurucz93,hauschildt99}.  However, in general this approach is often computationally prohibitive for most iterative methods of probabilistic inference. One partial compromise is to interpolate over a library of atmosphere structures that were pre-computed for a discrete set of parameter values, $\{\vt_{\ast}\}^{\rm grid}$, for some arbitrary $\vt_{\ast}$. Then, perform a radiative transfer calculation with that interpolated atmosphere to synthesize $\flam$ \citep[e.g., {\tt SME};][]{valenti96}.  A more common variant is to interpolate over a pre-synthesized library of model spectra, $\flam(\{\vt_{\ast}\}^{\rm grid})$ \citep[e.g.,][]{husser13,schonrich14}.  Although the former approach is preferable, the computational cost of repeated spectral synthesis is enough to make a detailed exploration of parameter space less appealing (although see Section~\ref{sec:discussion}).  Although the framework we are advocating is applicable for {\it any} ``back-end" that generates a model spectrum, it is illustrated here using the latter approach with the \citet{husser13} {\tt PHOENIX} library.

In practice, this reliance on spectral interpolation within a model library requires a sophisticated treatment of associated uncertainties.  The key problems are that the spectra themselves do not vary in a straightforward way as a function of $\vt_{\ast}$ (especially within spectral lines), and that the typical model library is only sparsely sampled in $\vt_{\ast}$.  Because of these issues, standard interpolation methods necessarily result in some information loss.  The practical consequence is that the inferred posteriors on the model parameters are often sharply peaked near a grid point in the library, $\{\vt_{\ast}\}^{\rm grid}$, potentially biasing the results and artificially shrinking the inferred parameter uncertainties (e.g., \citet{cottaar14}). 
To mitigate these effects, we develop a spectral ``emulator" that smoothly interpolates in a sparse model library and records a covariance term to be used in the likelihood calculation that accounts for the associated uncertainties.  The emulator is described in detail in Appendix~\ref{sec:Appendix}.  We first decompose the model library into a representative set of eigenspectra using a principal component analysis.  At each gridpoint in the library, the corresponding spectrum can be reconstructed with a  linear combination of these eigenspectra.  The weights associated with each eigenspectrum contribution vary smoothly as a function of the parameters, and so are used to train a Gaussian process to interpolate the weights associated with any arbitrary $\vt_{\ast}$.  In this way, the emulator delivers a probability distribution that represents the range of possible interpolated spectra.  By then marginalizing over this distribution, we can modify the likelihood function to propagate the associated interpolation uncertainty.  In the remainder of this section, the details of generating the reconstructed (interpolated) spectrum are not especially relevant (see Appendix~\ref{sec:Appendix}).

\subsection{Post-Processing} \label{subsec:postprocess}

Typically, the ``raw" interpolated model spectrum $\flam(\vt_{\ast})$ that was generated above is highly over-sampled, and does not account for several additional observational and instrumental effects that become important in comparisons with real data.  Therefore, a certain amount of post-processing is required before assessing the model quality.  We treat that post-processing in two stages.  The first stage deals with an additional set of ``extrinsic" parameters, $\vt_{\rm ext}$, that incorporate some dynamical considerations as well as observational effects related to geometry and the relative location of the target.  The second stage employs a suite of nuisance parameters, $\vp$, designed to forward model some imperfections in the data calibration.

\begin{figure}[!t]
\begin{center}
  \includegraphics{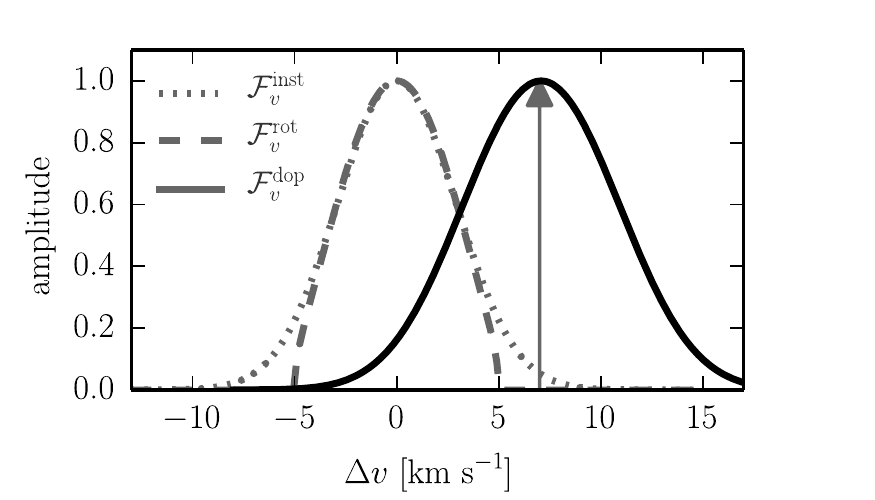}
  \includegraphics{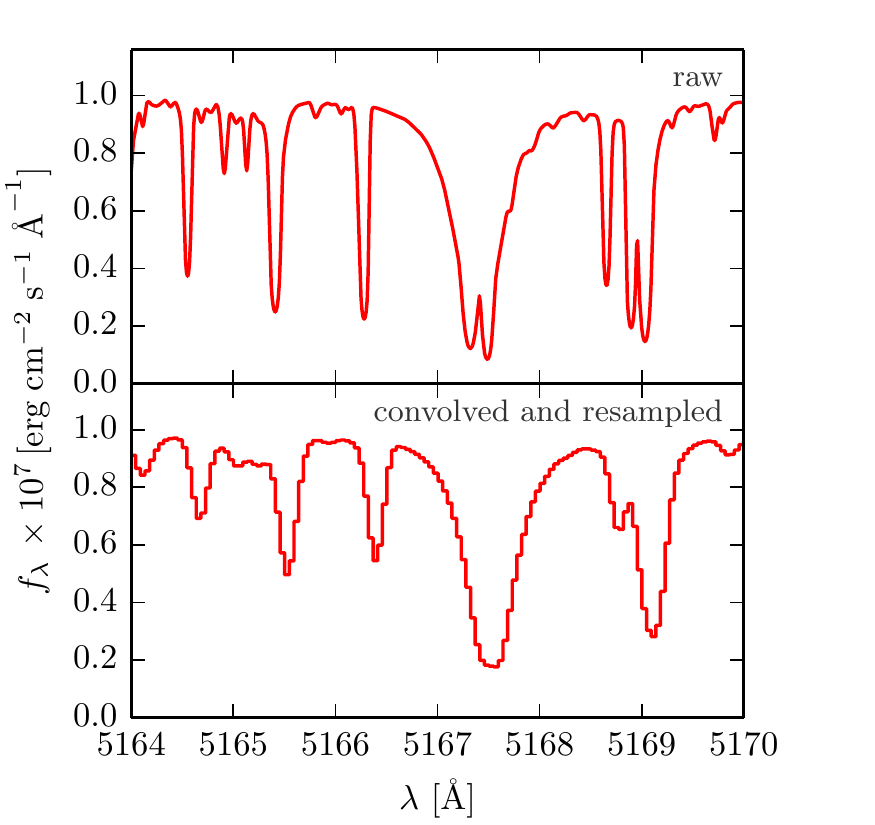}
  \figcaption{({\it top}) The line-of-sight velocity distribution function ({\it solid black curve}) and its decomposition into broadening kernels.  The instrumental kernel ({\it dotted}) is treated as a Gaussian, the rotation kernel ({\it dashed}) is a Gaussian-like function of the projected rotational velocity, and the Doppler kernel ({\it solid}) is a $\delta$-function that introduces the radial velocity.  In this specific case, $\sigma_v = 2.9$\,km s$^{-1}$, $\vsini = 5$\,km s$^{-1}$, and $v_r = 7$\,km s$^{-1}$, appropriate for the example in Section \ref{subsec:wasp}.  ({\it bottom}) A segment of a raw, full-resolution model spectrum and its post-processed equivalent after convolution and re-sampling at the coarser resolution of the detector pixels. 
\label{fig:broadening}}
\end{center}
\end{figure}

We can further divide $\vt_{\rm ext}$ into those parameters that impact the model primarily in the spectral or flux dimensions.  For the former, we consider three kernels that contribute to the line-of-sight velocity distribution function.  The first, $\mathcal{F}_v^{\rm inst}$, treats the instrumental spectral broadening. For illustrative purposes, we assume $\mathcal{F}_v^{\rm inst}$ is a Gaussian with a mean of zero and a constant width $\sigma_v$ at all $\lambda$, although more sophisticated forms could be adopted.  The second, $\mathcal{F}_v^{\rm rot}$, characterizes the broadening induced by stellar rotation, parameterized by $\vsini$ as described by \citet[][his Eq.~18.14]{gray08}, the rotation velocity at the stellar equator projected on the line of sight (where $i$ is the inclination of the stellar rotation axis).  And the third, $\mathcal{F}_v^{\rm dop} = \delta(v-v_r)$, incorporates the radial velocity through a Doppler shift.  The model spectrum is modified by the parameters $[\sigma_v, \,\, \vsini, \,\, v_r]$ through these kernels, using a convolution in velocity-space,\footnote{In practice, these convolutions are performed as multiplications in Fourier-space to better preserve spectral information \citep[cf.,][]{tonry79}; the mathematical formalism is presented for clarity.}
\begin{equation} \label{eqn:broadening}
\flam(\vt_{\ast}, \sigma_v, v\sin{i}, v_r) = \flam(\vt_{\ast}) \ast \mathcal{F}_v^{\rm inst} \ast \mathcal{F}_v^{\rm rot} \ast \mathcal{F}_v^{\rm dop}, 
\end{equation} 
and then re-sampled onto the discrete wavelengths corresponding to each data pixel, 
\begin{equation} \label{eqn:resampling}
\flam(\vt_{\ast},  \sigma_v, v\sin{i}, v_r) \mapsto \vM(\vt_{\ast},  \sigma_v, v\sin{i}, v_r),
\end{equation}
where the $\mapsto$ symbol denotes a re-sampling operator that maps the model spectrum onto the $N_{\rm pix}$-element model vector $\vM$ ($N_{\rm pix}$ is the number of pixels in the spectrum).  Figure \ref{fig:broadening} shows a (condensed) graphical representation of these post-processing steps.

At this stage, the model is further modified in the flux dimension.  A typical synthetic spectrum is computed as the flux that would be measured {\it at the stellar surface}, and so needs to be diluted by the subtended solid angle, $\Omega = (R_{\ast}/d)^2$, where $R_{\ast}$ is the stellar radius and $d$ is the distance.  An additional wavelength-dependent scaling factor is applied to account for interstellar extinction, assuming some previously-derived extinction law $A_{\lambda}$ \citep[e.g.,][]{cardelli89} that is parameterized by $A_V$.  The parameters $[\Omega, \,\, A_V]$ are then applied as
\begin{eqnarray} \label{eqn:scaling}
\vM(\vT) &=& \vM(\vt_{\ast}, \vt_{\rm ext}) \\
         &=& \vM(\vt_{\ast}, \sigma_v, v\sin{i}, v_r) \times \Omega \times 10^{-0.4\,A_{\lambda}}, \nonumber
\end{eqnarray}
with simplified notation such that $\vT \equiv [\vt_{\ast}, \,\, \vt_{\rm ext}]$, where $\vt_{\rm ext} = [\sigma_v, \vsini, v_r, \Omega, A_V]$. Some spectral libraries provide spectra as with peak fluxes normalized to a constant value, in that case, $\Omega$ will simply serve as an arbitrary scaling parameter.

The procedure so far is composed of straightforward operations demanded by practical astronomical and computing issues. If the data were {\it perfectly} calibrated, we could proceed to a likelihood calculation that makes a direct comparison with $\vM(\vT)$.  However, the calibration of the continuum shape for data with reasonably large spectral range is often not good enough to do this.  A common example of this imperfect calibration can be readily seen when comparing the overlaps between spectral orders from echelle observations.  Even if such imperfections (e.g., in the flat field or blaze corrections, or perhaps more likely in the flux calibration process) induce only minor, low-level deviations in the continuum shape, they can add up to a significant contribution in the likelihood function and thereby potentially bias the results.   

The traditional approach to dealing with this issue has been avoidance; a low-order polynomial or spline function is matched (separately) to the model and the data and then divided off to normalize the spectra.  While this is straightforward to do for earlier type stars, it only masks the problem.\footnote{For instance, the imperfect calibration would still in principle be discernible through the slight differences of the normalization functions derived for the data and model.}  This normalization procedure disposes of {\it useful} physical information content available in the continuum shape, and can be considerably uncertain in cases where the spectral line density is high (e.g., for cooler stellar photospheres).  Moreover, it can not propagate the uncertainty inherent in deriving the normalization functions into a proper inference framework.

\begin{figure}[t!]
\includegraphics{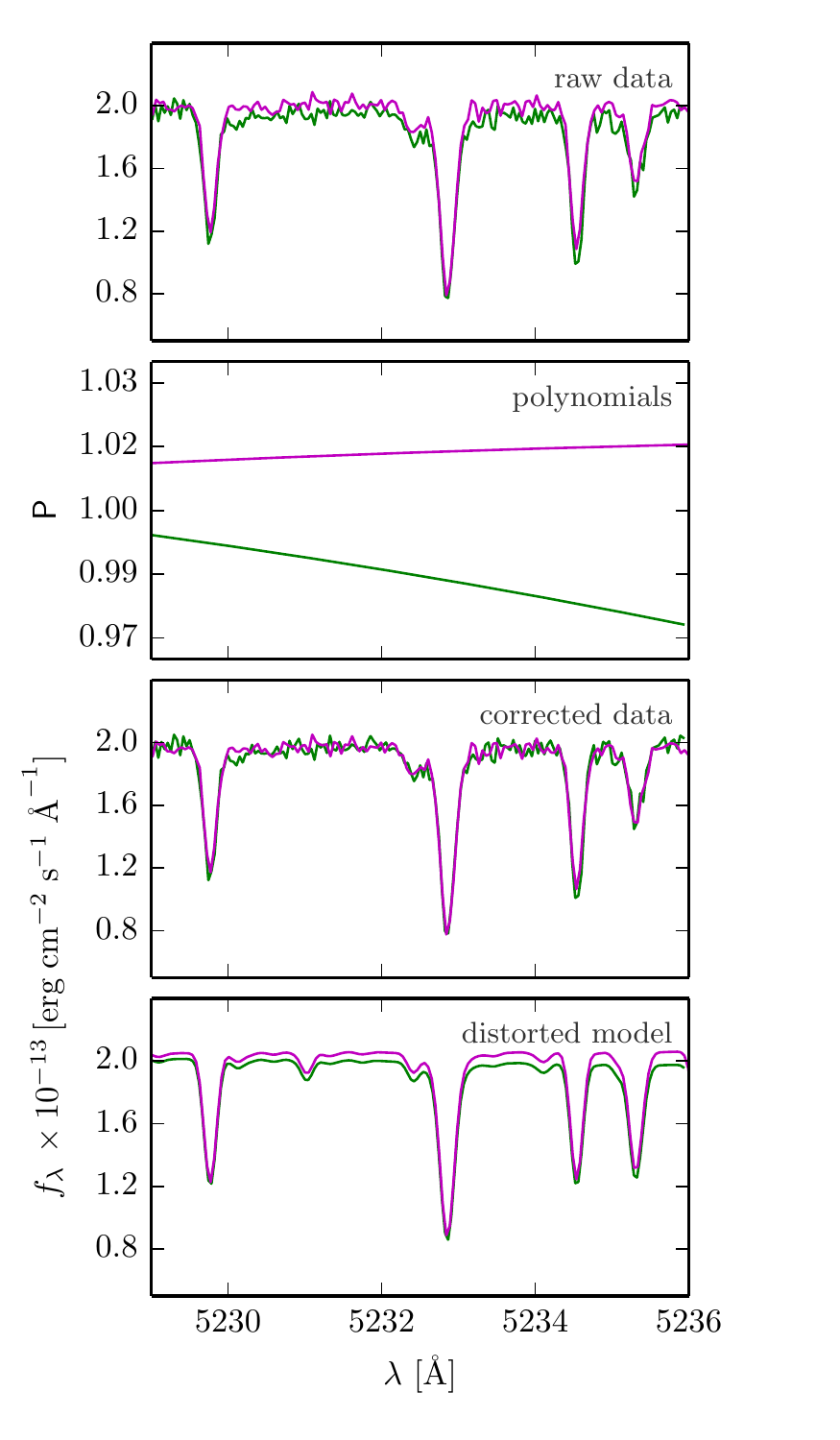}
\vspace{-0.75cm}
\figcaption{A demonstration of our treatment for residual calibration mismatches.  The observed spectra at the overlap of two echelle orders ({\it top}) have slightly ($\sim$1--3\%) discrepant continuum levels.  By using Chebyshev polynomials ({\it middle, top}) one can correct for that mismatch by adjusting the data ({\it middle, bottom}); instead, in practice we equivalently distort the model by these polynomials ({\it bottom}) such that the model remains linear in the Chebyshev coefficients (Eq.~\ref{eqn:chebyshev}). Note that this procedure preserves the natural units of flux and any intrinsic shape of the spectral energy distribution---the spectrum is \emph{not} continuum normalized. \label{fig:chebyshev}}
\end{figure}

Instead, we employ a more rigorous approach that forward-models the calibration imperfections with a set of nuisance parameters that modify the shape of the model spectrum.  By later marginalizing over these nuisance parameters, we properly account for any uncertainties that these kinds of calibration imperfections induce on the stellar parameters of interest while also maintaining the useful information in the continuum shape.  In practice, this is achieved by distorting segments of the model with polynomials, $\mathsf{P}$ \citep[e.g.,][]{eisenstein06,koleva09}. Figure \ref{fig:chebyshev} demonstrates how these nuisance parameters are applied to the model. For $N_{\rm ord}$ spectral orders, each denoted with index $o$, the model spectrum can be decomposed as
\begin{eqnarray} \label{eqn:chebyshev}
\vM(\vT, \cheb) &=& \{ \vM_o(\vT) \times \mathsf{P}_o \} \\
                &=& \{ \vM_o(\vT) \times \sum_n c_o^{(n)} \, T_o^{(n)} \}, \nonumber
\end{eqnarray}
where $T^{(n)}$ is an $n^{\rm th}$ degree Chebyshev function. 
The $n N_{\rm ord}$ coefficients are considered a set of nuisance parameters, $\cheb = [\{c_o^{(0)}, c_o^{(1)}, \ldots, c_o^{(n-1)} \}]$.  Judicious priors can ensure that the real spectral features (e.g., molecular bands) are not treated as residual calibration artifacts.  The lowest-degree (scaling) coefficient, $c^{(0)}$, is degenerate with the solid angle, $\Omega$.  Therefore, we enforce an additional constraint that the mean of the polynomial is unity.  For data with a single spectral order, this means simply setting $c^{(0)} = 1$.  In the multiple order case, we assign $c^{(0)} = 1$ in an arbitrary order as an anchor, but permit the $c^{(0)}$ in other orders to be different.   

\subsection{Model Evaluation} \label{subsec:likelihood}

The fit of the model spectrum is assessed by comparing to the data with a pixel-by-pixel likelihood calculation.  If we denote the data spectrum as $\vD$, then a corresponding residual spectrum (an $N_{\rm pix}$-element vector) can be defined for any input parameter set,
\begin{equation}
\vR \equiv \vR(\vT, \cheb) \equiv \vD-\vM(\vT, \cheb).
\end{equation}
To quantify the probability of the data conditioned on the model, we adopt a standard 
multi-dimensional Gaussian likelihood function
\begin{equation}
p(\vD|\vM) =  \frac{1}{[(2 \pi)^{N_{\rm pix}} \det(\vC)]^{1/2}} \exp\left( -\frac{1}{2}
   \vR^\trans \vC^{-1} \vR \right)
   \label{eqn:likelihood}
\end{equation}
that penalizes models which yield larger residuals and explicitly allows for covariances in the residual spectrum through the $N_{\rm pix} \times N_{\rm pix}$ matrix $\vC$.  For practical reasons, the log-likelihood is used as the quality metric, where
\begin{equation}
  \ln{p(\vD | \vM)} = -\frac{1}{2} \left( \vR^\trans \vC^{-1} \vR + \ln{\det{\vC}} + N_{\rm pix} \ln{2\pi} \right).
  \label{eqn:lnlikelihood}
\end{equation}

The covariance matrix $\vC$ characterizes both the measurement uncertainty ($\sigma$; ``noise") in each pixel and the covariance between pixels.  When using a spectral emulator to interpolate model spectra, $\vC$ will be the sum of the covariance matrix described here and the emulator matrix derived in Appendix~\ref{sec:Appendix} (Eq.~\ref{eqn:modC}).  In the special case where each pixel is independent, the covariance matrix is diagonal, $\vC_{ij} = \delta_{ij} \,\, \sigma_i^2$, where $\sigma_i$ is the uncertainty in pixel $i$ and $\delta_{ij}$ is the Kronecker delta function, and Eq.~\ref{eqn:lnlikelihood} reduces to the familiar
\begin{equation}
  \ln{p(\vD | \vM)} - \textrm{constant} = -\frac{1}{2} \sum_i^{N_{\rm pix}} \frac{\vR_i^2}{\sigma_i^2} \equiv -\frac{\chi^2}{2},
\label{eqn:chisq}
\end{equation}
the sum of the square of the residuals weighted by their inverse variances.  However, that simplification rarely applies in practice.  A more complex covariance matrix is required, so that additional off-diagonal terms can be used to explicitly characterize (1) pixel-to-pixel covariances imposed by the discrete over-sampling of the line-spread function, and (2) highly correlated residuals as manifestations of systematic imperfections in the model library.  The following sections describe how these issues are addressed by constructing a more sophisticated $\vC$.

\subsubsection{Global Covariance Structure} \label{subsec:global_covariance}

\begin{figure}[!t]
\begin{center}
  \includegraphics{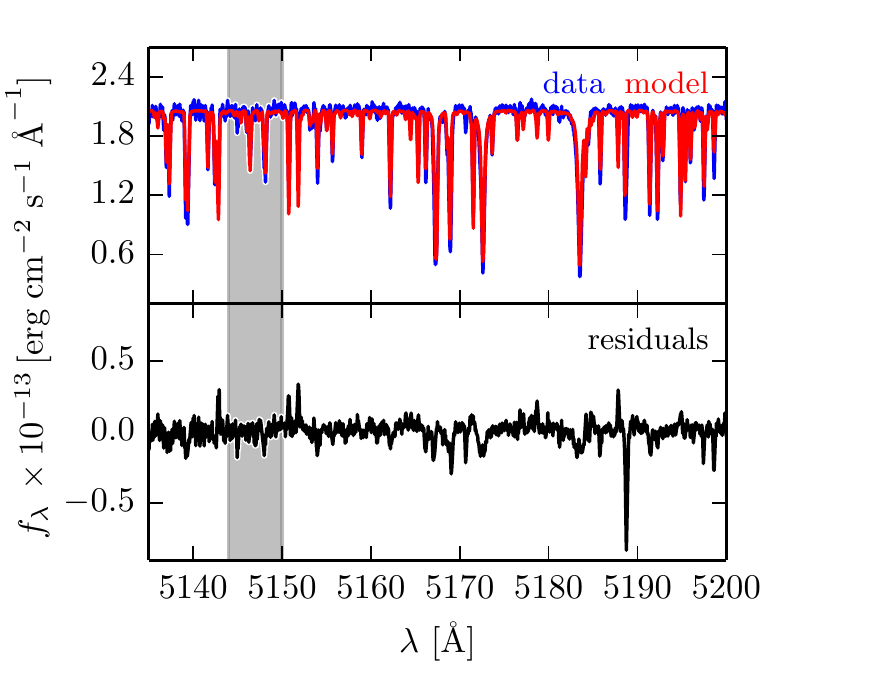}
  \includegraphics{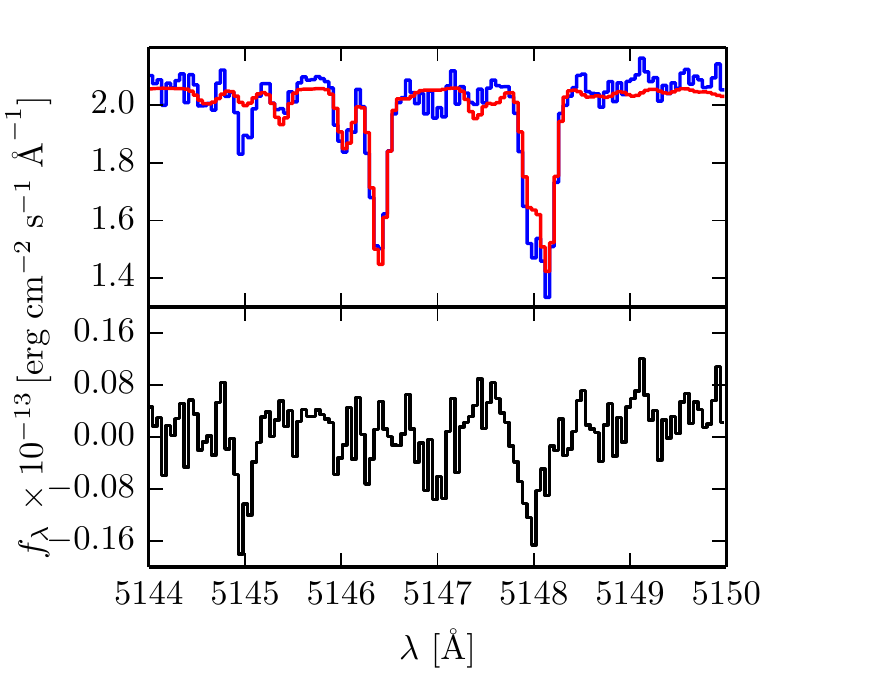}
  \includegraphics{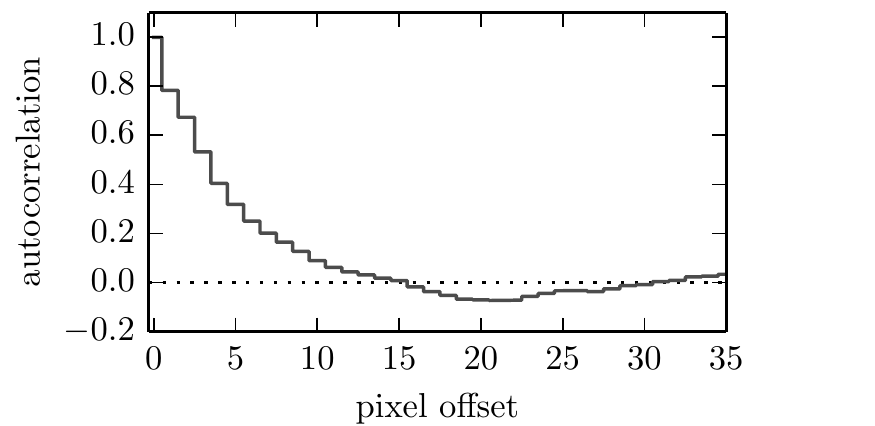}
  \figcaption{({\it top}) A comparison of the data and a typical model with parameters drawn from the posterior distribution, along with the corresponding residual spectrum.  ({\it middle}) A zoomed view of the gray band in the top panels, highlighting the mildly covariant residual structure that is produced by slight mismatches between the data and model spectra.  ({\it bottom}) The autocorrelation of the residual spectrum.  Notice the substantial autocorrelation signal for offsets of $\lesssim 8$ pixels, demonstrating clearly that the residuals are not well described by white (Poisson) noise alone. \label{fig:class0}}
\end{center}
\end{figure}

Astronomical spectrographs are designed to have the detector over-sample the instrumental line-spread function with at least a few pixels.  Therefore, adjacent pixels never record independent samples of the true spectrum.  In that case, a difference between an observed and modeled spectral feature creates a correlated residual that spans multiple pixels.  This can be demonstrated clearly in the autocorrelation of $\vR$: a slight model mismatch will produce correlated residuals over a characteristic scale similar to the instrumental or rotation broadening kernel width (whichever is larger).  Figure \ref{fig:class0} shows an example of these correlated residuals in real data; a significant autocorrelation signal is seen on an $\sim$8 pixel scale, corresponding to the 6.8\,km s$^{-1}$ FWHM of $\mathcal{F}_v^{\rm inst}$.  

It is important to distinguish here between ``noise" and the fit residuals.  Noise introduced to the spectrograph by astrophysical or instrumental effects is generally uncorrelated with wavelength.  The arrival and propagation of each photon through the instrument and into the detector can be considered an independent event.  In essence, the noise itself is not correlated, but the fit residuals likely are.  However, from a mathematical perspective the correlated residuals can be treated in the same way as correlated noise, by constructing a non-trivial covariance matrix with off-diagonal terms.  In practice, this is achieved by parameterizing $\vC$ with a kernel that describes the covariance between any pair of pixels, indexed $ij$, representing wavelengths $\lambda_i$ and $\lambda_j$.

\begin{figure*}[!htb]
\begin{center}
\includegraphics[scale=1.05]{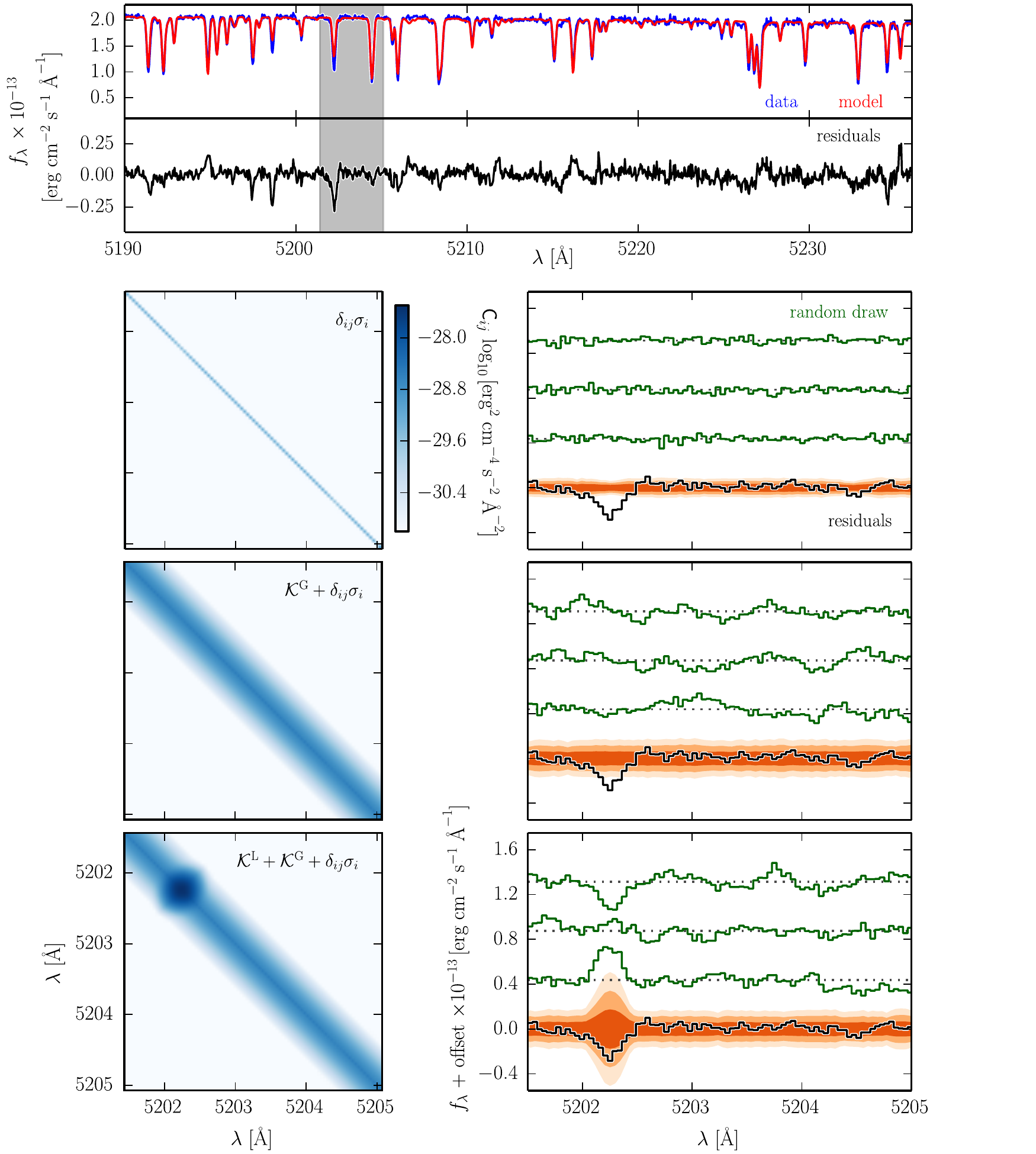}
\figcaption{A decomposition of the modeling procedure, explicitly highlighting the roles of the various contributions to the covariance matrix.  The top panels show a typical comparison between the data and model spectrum, along with the associated residual spectrum.  The subsequent panels focus on the illustrative region shaded in grey.  The left column of panels show the corresponding region of the covariance matrix $\vC$, decomposed into its primary contributions: from top to bottom, the trivial noise matrix, then combined with the global covariance kernel, and finally including an appropriate local covariance kernel.  In the right column, we show the zoomed-in residual spectrum ({\it black}) along with example random draws from the subsets of $\vC$ exhibited to the left.  The shaded contours ({\it orange}) represent the 1, 2, and 3\,$\sigma$ dispersions of an ensemble of 200 random draws from $\vC$.  Note that the trivial noise matrix ($\delta_{ij} \sigma_i$) poorly reproduces both the scale and structure of the residual spectrum.  The addition of a global kernel ($\mathcal{K}^\textrm{G}$) more closely approximates the structure and amplitude of the residuals, but misses the outlier line at 5202.2\,\AA.  Including a local kernel ($\mathcal{K}^\textrm{L}$) at that location results in a covariance structure that does an excellent job reproducing all the key residual features.  \label{fig:matrix}}
\end{center}
\end{figure*}

For a well-designed spectrograph and sufficiently accurate model, this {\it global} (i.e., present throughout the spectrum) covariance should have a relatively low amplitude and small correlation length.  To describe that structure, we use a stationary covariance kernel (or radial basis function) with an amplitude that depends only on the velocity separation between two pixels,
\begin{equation}
  r_{ij} \equiv r(\lambda_i, \lambda_j) = \frac{c}{2} \left | \frac{\lambda_i 
   - \lambda_j}{ \lambda_i + \lambda_j} \right |,
\end{equation}
where $c$ is the speed of light. This kernel is used to characterize the covariance between pixel residuals, 
\begin{equation}
  \Kglobal_{ij} =  \langle \vR_i \; \vR_j \rangle.
  \label{eqn:expectation}
\end{equation}
A variety of useful kernels have been developed in the field of Gaussian processes to parameterize such a covariant structure \citep[e.g.,][]{rasmussen05,santner13}, and are seeing increased use in many areas of astrophysics \citep[for some specific examples in stellar and planetary applications, see][]{foreman-mackey14,aigrain15,barclay15}.  After some experimentation, we adopted the \matern\ kernel with $\nu = 3/2$ because it performed well at reproducing the appearance of realistic residuals for this specific problem.  In this case,   
\begin{equation}
  \Kglobal_{ij}(\vp_{{\mathsf C}, {\rm G}}) = w_{ij} a_{\rm G} \left(1 + \frac{\sqrt{3}\, r_{ij}}{\ell} \right ) \exp 
   \left (- \frac{\sqrt{3}\, r_{ij}}{\ell} \right ),
   \label{eqn:global}
\end{equation}
with $\vp_{{\mathsf C}, {\rm G}} = [a_{\rm G}, \ell]$, an amplitude ($a_{\rm G}$) and a scale ($\ell$). The $\vp_{{\mathsf C}, {\rm G}}$ are termed {\it hyperparameters} here; because a Gaussian process describes a population of functions generated by random draws from a probability distribution set by a mean vector and a covariance matrix, the kernel parameters are naturally part of a hierarchical model. In this specific case, the functions described by these hyperparameters  represent many realizations of covariant residuals from a spectral fit. Figure~\ref{fig:matrix} shows an example of the Gaussian process kernel and the covariant residuals that can be generated from it.  To ensure that $\vC$ remains a relatively sparse matrix (for computational expediency), we employ a Hann window function
\begin{equation}
  w_{ij} \,(r_0) = \left \{ 
    \begin{array}{cc}
    \frac{1}{2} + \frac{1}{2} \cos \left(\frac{\pi r_{ij}}{r_0} \right) & r_{ij} \le r_0 \\
    0 & r_{ij} > r_0 \\
  \end{array}
  \right .
  \label{eqn:Hann}
\end{equation}
to taper the kernel.  The truncation distance $r_0$ can be set to a multiple of the scale (we set $r_0 = 4\ell$).  

\subsubsection{Local Covariance Structure} \label{subsec:local_covariance}

\begin{figure}[!t]
\begin{center}
\includegraphics{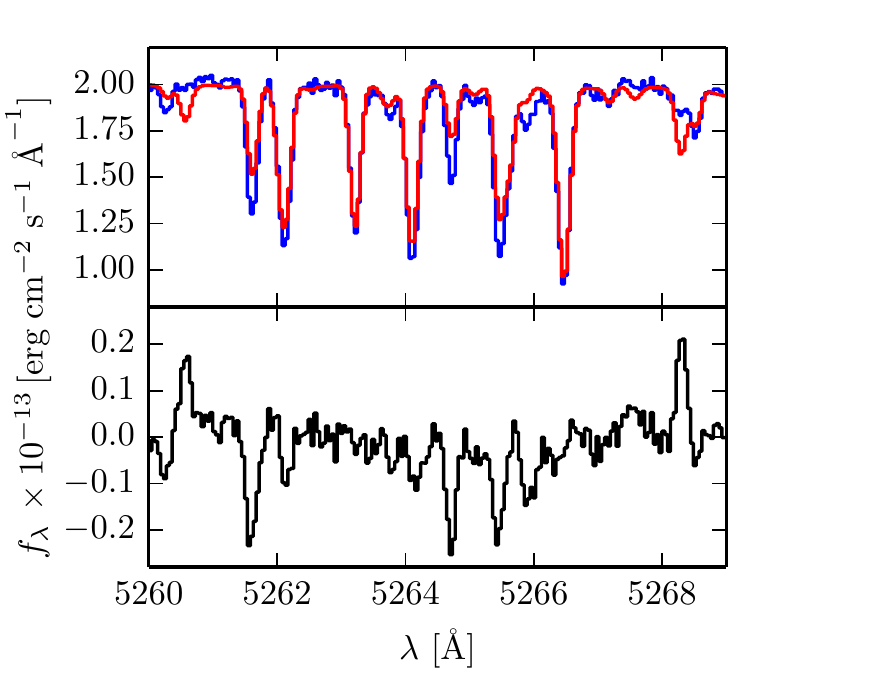}
\figcaption{A particularly illustrative spectral region with substantial localized structure in the residuals due to ``outlier" spectral lines in the model library.  For any specific line, there might exist a set of model parameters, $\vT$, that will improve its match with the data, but a $\vT$ that will properly fit \emph{all} the outlier lines does not exist in a pre-computed library with (necessarily) limited parametric flexibility.  Out of concern that such intrinsic mismatches can bias the inference on $\vT$, the methodology advocated here introduces local kernels to inflate the covariance around these outliers, self-consistently down-weighting their influence on the fit.  
\label{fig:badlines}}
\end{center}
\end{figure}

In addition to the global covariance structure, there can be local regions of highly correlated residuals.  These patches of large $\vR$ are usually produced by pathologically incorrect spectral features in the model, due to systematic imperfections like missing opacity sources or poorly constrained atomic/molecular data (e.g., oscillator strengths).  Some representative examples are shown in Figure~\ref{fig:badlines}.  To parameterize such regions in $\vC$, we introduce a sequence of non-stationary kernels that explicitly depend on the actual wavelength values of a pair of pixels ($\lambda_i$ and $\lambda_j$), and not simply their separation ($r_{ij}$).  

Assuming that these local residual features are primarily due to discrepancies in the spectral line depth (rather than the line shape or central wavelength), a simple Gaussian is a reasonable residual model. In that case, the pixel residuals of the $k$-th such local feature could be described as
\begin{equation}
\vR_j  \equiv \vR(\lambda_j) = A_k \exp \left [ - \frac{r^2(\lambda_j, \mu_k)}{2 \sigma_k^2} \right ]
\end{equation}
with peak amplitude $A_k$, central wavelength $\mu_k$, and width $\sigma_k$. We assume that the amplitude of this Gaussian feature is drawn from a normal distribution
\begin{equation}
A_k \sim {\cal N} ( 0, a_k^2)
\label{eqn:amplitude}
\end{equation}
with mean 0 and variance $a_k^2$. The pixels in this Gaussian-shaped residual are correlated because each pixel shares a common random scale factor ($A_k$). Then, the covariance of any two pixels in this region is given by Eq.~\ref{eqn:expectation}, where the expectation value is taken with respect to the probability distribution in Eq.~\ref{eqn:amplitude}
\begin{eqnarray}
\mathcal{K}^{\textrm{L},k}_{ij} &=& \left \langle A_k \exp \left [ - \frac{r^2(\lambda_i, \mu_k)}{2 \sigma_k^2} \right ]  A_k \exp \left [ - \frac{r^2(\lambda_j, \mu_k)}{2 \sigma_k^2} \right ] \right \rangle \nonumber \\
&=& \langle A_k^2 \rangle \exp \left [ - \frac{r^2(\lambda_i, \mu_k) + r^2(\lambda_j, \mu_k)}{2 \sigma_k^2} \right ] \nonumber \\
&=& a_k^2 \exp \left [ - \, \frac{r^2(\lambda_i, \mu_k) + r^2(\lambda_j, \mu_k)}{2 \sigma_k^2}\right ]. 
\label{eqn:kregion}
\end{eqnarray}
The full local covariance kernel covering all of the possible Gaussian residuals is composed of a linear combination of kernels,
\begin{equation} \label{eqn:klocal}
  \Klocal_{ij}(\vp_{{\mathsf C},L}) = \sum_k^N w^k_{ij} \, \mathcal{K}^{\textrm{L},k}_{ij},
\end{equation}
with a corresponding set of hyperparameters $\vp_{{\mathsf C},\textrm{L}} = [\{a_1, \mu_1, \sigma_1, \ldots, a_N, \mu_N, \sigma_N\}]$.  Note that we again taper the kernels with Hann windows (Eq.~\ref{eqn:Hann}) to ensure a sparse covariance matrix; in this case, the truncation distance $r_0$ can be set to some multiple of the width parameter (e.g., $r_0 = 4\sigma_k$).  In effect, these kernels systematically down-weight the influence of strong residuals in the likelihood calculation, mitigating any potential bias they might induce on inferences of the interesting parameters ($\vT$).  Similar in spirit to robust linear regression and ``bad data" mixture models \citep{hogg10}, these kernels provide a means for (correlated) outlier rejection that preserves the integrity of the probabilistic framework (as opposed to the common manual or threshold-based techniques of masking or clipping). 

In principle, the concept of these local kernels can be extended to account for more complex residual structures.  For example, late-type stars with imperfectly modeled molecular bandheads may produce a complicated pattern of positive and negative residuals or a pronounced mismatch over a relatively large spectral scale.  This phenomenologically different local covariance behavior can still be treated in this framework if an appropriate kernel morphology is adopted.

\subsubsection{Composite Covariance Matrix}

We can now compute the covariance matrix employed in the likelihood calculation (Eq.~\ref{eqn:lnlikelihood}) as the linear combination of the trivial pixel-by-pixel noise matrix and the global and local kernels discussed above, 
\begin{equation}
\vC_{ij}(\cov)  = b \, \delta_{ij} \, \sigma_i^2 + \Kglobal_{ij}(\vp_{{\mathsf C}, G}) + 
                  \Klocal_{ij}(\vp_{{\mathsf C}, L}), 
\end{equation}
with hyperparameters $\cov = [\vp_{{\mathsf C}, \textrm{G}}, \vp_{{\mathsf C}, \textrm{L}}]$.  The factor $b$ is a parameter that scales up the Poisson noise in each pixel by a constant factor to account for additional detector or data reduction uncertainties (e.g., read noise, uncertainties in the spectral extraction procedure, etc.); typically $b < 1.1$ for well-calibrated optical spectra.  If there are $N_{\rm loc}$ local covariance patches (see Section \ref{subsec:MCMC} on how this is determined), then there are $4N_{\rm loc}+2$ elements in the set of covariance hyperparameters, $\cov$.  Figure \ref{fig:matrix} provides a graphical illustration of how the kernels that comprise the covariance matrix are able to reproduce the structure present in a typical residual spectrum. 

\subsection{Priors} \label{subsec:priors}

The Bayesian framework of this inference approach permits us to specify prior knowledge about the model parameters, $p(\vM)$.  As will be discussed further in Sections~\ref{sec:examples} and \ref{sec:discussion}, in most cases it is necessary to utilize some independent information (e.g., from asteroseismology constraints or stellar evolution models) as a prior on the surface gravity.  But otherwise we generally recommend a conservative assignment of uniform priors, such that $p(\vt_{\ast})$ is flat over the spectral library grid (and zero elsewhere) and $p(\vt_{\rm ext})$ is flat for physically meaningful values (e.g. $\vsini \ge 0$, $\Omega > 0$, and $A_V \ge 0$). 

\begin{figure}[!b]
  \includegraphics{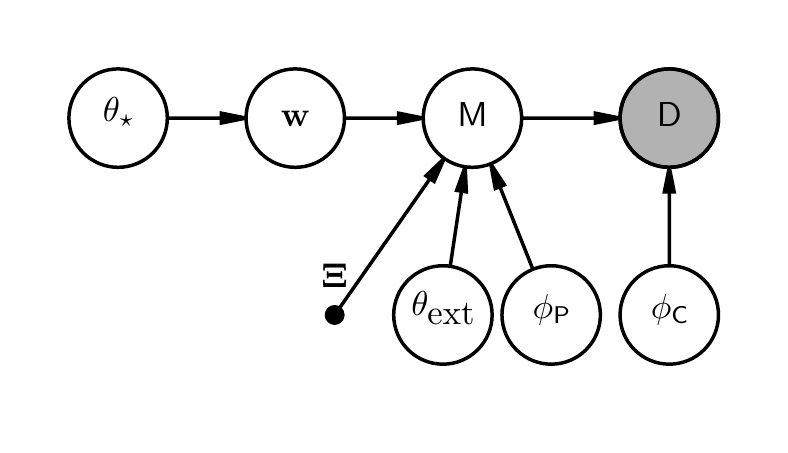}
  \figcaption{A probabilistic graphical model representing how the parameters of the model are    combined to forward model a spectrum and evaluate the likelihood function (Eq.~\ref{eqn:likelihood}).  When interpolating models using a spectral emulator (Appendix \ref{sec:Appendix}), the stellar parameters ($\vt_\ast$) determine the weights    ($\mathbf{w}$) of the eigenspectra ($\mathbf{\Xi}$), which are modified according to the    observational parameters ($\vt_\textrm{ext}$) and polynomial parameters ($\cheb$). Together,    these parameters specify the model spectrum ($\vM$). If one uses linear interpolation instead of a spectral emulator, then there would be no intermediate nodes for $\mathbf{w}$ and $\mathbf{\Xi}$. The structure of the covariance matrix, which is included in the likelihood function, is determined by the covariance hyperparameters ($\cov$). Together, the model spectrum and the covariance matrix predict the resulting dataset ($\vD$).
\label{fig:PGM}}
\end{figure}

For (early type) stars with a clear continuum, it makes sense to assume flat priors on the polynomial parameters $\cheb$.  However, information about the calibration accuracy (e.g., from comparisons of multiple calibration sources in the same observation sequence) can be encoded into a simple prior on the Chebyshev coefficients; for example, Gaussian priors with widths that represent the fractional variance between different derived calibration functions would be reasonable.  For (late type) stars with a poorly defined continuum, some judicious tapering of the priors (such that small coefficients at high $n$ are preferred) may be required to ensure that broad spectral features are not absorbed into the polynomial (see Section~\ref{sec:examples}). 

In general, uniform (non-negative) priors are recommended for the global kernel hyperparameters.  For the local kernels, we typically adopt uniform priors for the amplitudes and means \{$a_k$, $\mu_k$\}, but construct a logistic prior for the widths \{$\sigma_k$\} that is flat below the width of the line-of-sight velocity distribution function (defined as the convolution of three broadening kernels in Eq.~\ref{eqn:broadening}), $\sigma_{\rm los}$, and smoothly tapers to zero at larger values:
\begin{equation}
p(\sigma_k) = \frac{1}{1 + e^{\,\sigma_k-\sigma_{\rm los}}}.
\end{equation}
Such a prior formulation prevents local kernels from diffusing to large $\sigma_k$ and low $a_k$, since that kind of behavior is better treated by the global kernel. When modeling real data, there is no \emph{a priori} information about the locations $\{\mu_k \}$ of the local kernels; they are instantiated as needed (see Section~\ref{subsec:MCMC}). However, using the knowledge gained from previous inferences of similar targets, one could instead start by instantiating kernels at the outset with priors on $\{\mu_k \}$ where there are known to be systematic issues with the synthetic spectra.

\subsection{Exploring the Posterior} \label{subsec:MCMC}

The inference framework developed here has a natural blocked structure between the collections of ``interesting" parameters, $\vT = [\vt_{\ast}, \vt_{\rm ext}]$, the nuisance parameters $\cheb$, and the covariance hyperparameters $\cov$. 
The conditional dependencies of these parameters are shown graphically in Figure~\ref{fig:PGM} as a directed acyclic graph \citep{bishop06,mandel09}. To explore the posterior distribution, 
\begin{equation} 
p(\vT, \cheb, \cov | \vD) \propto p(\vD | \vT, \cheb, \cov) \, p(\vT, \cheb, \cov)
\label{eqn:post}
\end{equation}
for this type of structure, it is convenient to employ Markov Chain Monte Carlo (MCMC) simulations with a blocked Gibbs sampler coupled to the Metropolis-Hastings algorithm.  This procedure works by sampling in a subset of parameters (with Metropolis-Hastings proposals) conditioned on the current (fixed) values of the other parameters.  After each iteration, the Gibbs sampler updates the sampled parameters and then cycles through all the (previously fixed) different parameter subsets in the same way (for a more mathematical description, see Chapter 11 of \citealt{gelman13}).  A step-by-step prescription follows, where the $i^{\rm th}$ iteration of the Gibbs sampler is indexed with a superscript: \\

\noindent (1) Initialize the parameters.  One might set $\vT^0$ based on estimates in the literature or scaling behaviors, and make simple assumptions about $\cheb^0$.  Here, we set the Chebyshev coefficients ($\cheb^0$) so that the polynomials are constant ($c_o^{(0)} = 1$ and $c_o^{(>0)} = 0$, $\forall$ $o$) and assume only the trivial noise spectrum (and spectral emulator kernel; see Appendix~\ref{sec:Appendix}) contributes to the $\vC$ (i.e., $\cov^0 = 0$).  \\

\noindent (2a) Start the $i^{\rm th}$ iteration of the Gibbs sampler.  For each iteration of the Metropolis-Hastings algorithm, sample in $\vT$ to evaluate the posterior (Eq.~\ref{eqn:post}) following the framework laid out in Sections \ref{subsec:likelihood} and \ref{subsec:priors}.  This represents a ``slice" through the posterior space conditioned on the other parameters being held fixed ($\cheb = \cheb^{i-1}$ and $\cov = \cov^{i-1}$).  Then update $\vT^{i-1} \rightarrow \vT^i$. \\

\noindent (2b) For each spectral order, sample in the polynomial parameters $\cheb$ and covariance hyperparameters $\cov$, conditioned on the other parameters being held fixed $\vT^i = \vT^{i-1}$.  Then update $\cheb^{i-1} \rightarrow \cheb^i$ and $\cov^{i-1} \rightarrow \cov^i$. \\

\noindent (3) Repeat Step~(2) for 20,000 samples. \\

\noindent (4) Repeat the procedure in Steps~(1)--(3) with different initializations, storing the samples for each Markov chain. After removing the burn-in samples for each chain, we compute the Gelman-Rubin convergence diagnostic, $\hat{R}$ \citep[][their Eq.~11.4]{gelman13}. If $\hat{R} \, < \, 1.1$, we can be reasonably sure that all of the chains have converged to the posterior distribution. \\

In Step~(2b), local covariance kernels are instantiated according to the following procedure.  First, an ``average" residual spectrum is generated by combining $\sim$500 residual spectra that were stored during a burn-in period using only the global kernels (prior to this storage, the Markov chain is thinned to account for autocorrelation of the posterior samples).  This average spectrum is then iteratively examined for deviations outside a critical threshold.  When a large residual is identified, a local kernel is introduced with a mean ($\mu_k$) at its location.\footnote{Although this may seem similar to the procedure of ``sigma-clipping", there is a crucial difference.  Rather than rejecting outlier data once it is found (i.e., setting its weight in the inference problem to zero), this procedure will actually self-consistently determine how to weight the outliers inside the probabilistic framework.}  After some experimentation with different threshold criteria, we chose to instantiate when the local residual is $>$4$\times$ the standard deviation in the average residual spectrum.\footnote{Lower thresholds result in more local kernels, thereby reducing the amplitude of the global kernel.  In the extreme case of a very low threshold, a local kernel would be instantiated for every spectral line (and no global kernel would be required).  We found that ultimately the posterior inferences on the parameters of interest are relatively insensitive to the choice of a threshold level, so long as it is set low enough to capture the egregious outliers.}  Alternative schemes, such as re-evaluating the kernel locations with each iteration of the Gibbs sampler, yield similar results; however, the adopted approach consistently converges with minimal computational overhead.  Once all the local kernels are instantiated, the Gibbs sampler is run for another period of burn-in.\footnote{There is no practical reason to delete local kernels once instantiated.  If the parameters have changed such that a given local kernel is no longer required, that kernel amplitude will be driven towards zero and represent a negligible contribution to $\vC$; in effect, the model will act as if the kernel were deleted automatically.}

This entire procedure can be a significant computational challenge.  A typical spectrum has $N_{\rm pix} > \mathcal{O}(10^3)$, and therefore the many evaluations of the matrix product $\vR^{\trans} \vC^{-1} \vR$ in the likelihood calculation can be numerically expensive.  However, because $\vC$ is a symmetric, positive semi-definite matrix, we can employ Cholesky factorization to optimize the evaluation of the matrix product and avoid the direct calculation of the matrix inversion ($\vC^{-1}$).  For multi-order echelle spectra or multiple spectra of the same target (perhaps taken with different instruments), the nuisance parameters for each segment of the spectrum are independent. This means that the computationally intensive steps of generating a model spectrum for a specific wavelength range and evaluating the likelihood can be parallelized. The only segment of the code that needs to be synchronized is the MCMC proposal of stellar parameters, which are shared between all chunks of the spectrum.  The massive parallelization of this algorithm on a computer cluster therefore enables the simultaneous inference of interesting parameters over wide spectral ranges at high resolution, or from multiple datasets.  To sample the posteriors in this mode, we extend the Metropolis-Hastings sampler included in the {\tt emcee} package \citep{foreman-mackey13} to function within a parallelized blocked Gibbs sampler.

The time required to thoroughly explore the posterior depends on both the data volume and the desired precision on the inference of the covariance hyperparameters.  If only the stellar parameters $\vT$ are of interest, one can first optimize the kernel parameters and then proceed with them fixed, since the stellar parameter posteriors are relatively insensitive to the precise value of the kernel parameters (once near their optimal value).  A fit of an $R\approx40,000$ spectrum with $>$30 echelle orders takes $\sim$2 hours (parallelized on a cluster).  If the full posteriors for the nuisance parameters are desired, the computation might take an order of magnitude longer.

\section{Demonstrations} \label{sec:examples}

In this section, we illustrate how the modeling framework operates for two real datasets.  The first is an elaboration of the example shown throughout Section \ref{sec:method}, using a high resolution optical spectrum of the F5 star (and transiting exoplanet host) WASP-14 \citep{joshi09,torres12}.  The second uses a medium resolution near-infrared spectrum of the M5 dwarf Gliese 51 (hereafter Gl 51), observed as part of the NASA/IRTF library of spectral standards \citep{cushing05,rayner09}.    
In both cases, we sequentially build up the complexity of the modeling framework to demonstrate how each of the components described in Section~\ref{sec:method} affects the posteriors on the parameters of interest ($\vt_{\ast}$).  We adopt the recent incarnation of the {\sc Phoenix} library \citep{husser13} for the models, although comment on systematic differences between libraries in Section~\ref{sec:systematics}.

\subsection{WASP-14} \label{subsec:wasp}

A high resolution ($R\approx44,000$) optical spectrum of WASP-14 was obtained on 2009 June 14 using the Tillinghast Reflector Echelle Spectrograph \citep[TRES;][]{furesz08} on the Fred Lawrence Whipple Observatory 1.5\,m telescope.  TRES delivers an echelle spectrum with 51 orders that cover the full optical range, from 3860--9100\,\AA.  The data were reduced and calibrated using standard techniques in the TRES pipeline (cf.,~\citealt{buchhave10}; see \citealt{torres12} for more specific details).  At 5100\,\AA, the S/N is $\sim$150 per resolution element.  Following \citet{torres12}, we focus here on the central three TRES orders, covering $\sim$5100-5400\,\AA.

\begin{figure}[!b]
  \includegraphics[width=0.5\textwidth]{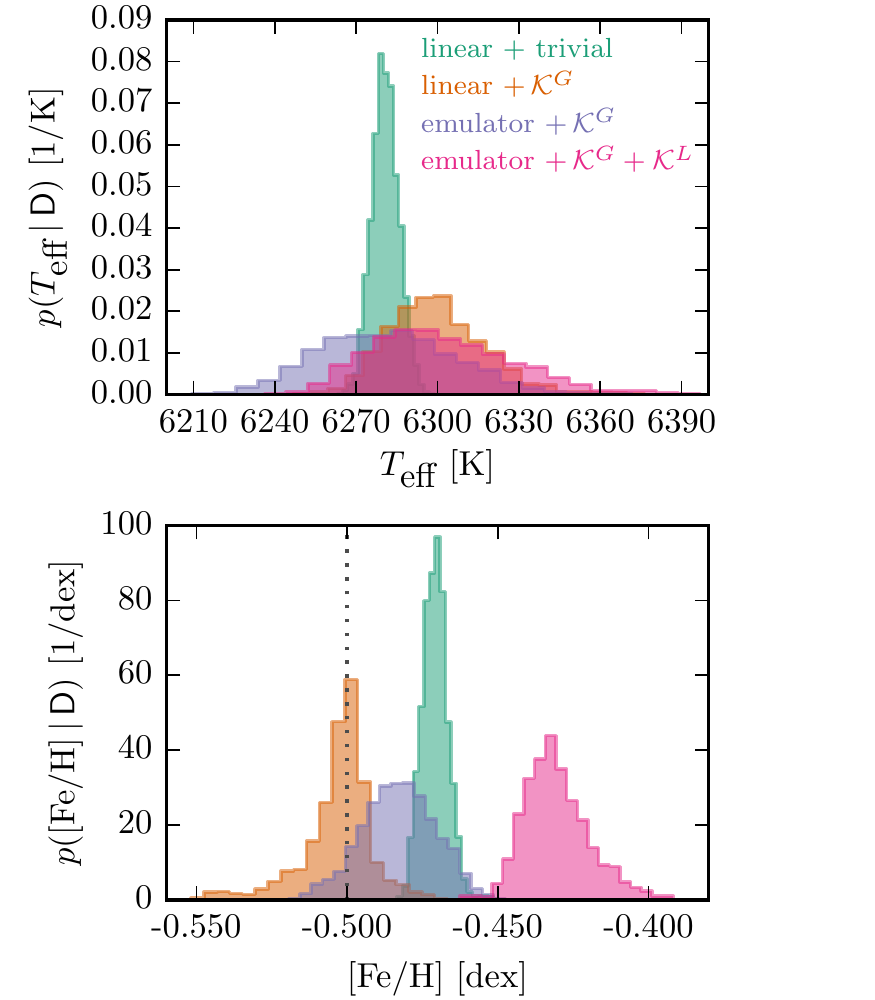}
\vspace{-0.5cm}
  \figcaption{The marginal posterior probability distributions for the WASP-14 $\teff$ and $\Z$ based on the {\sc PHOENIX} model library, for various levels of model complexity, including: (1) a simple linear interpolation scheme and trivial covariance matrix ({\it blue-green}); (2) including global covariance terms from Gaussian process kernels ({\it orange}); (3) employing a Bayesian emulator for more appropriate interpolation ({\it purple}); and (4) also including local covariance kernels to downweight systematic outlier spectral lines ({\it magenta}).  \label{fig:PHOENIX_posterior} }
\end{figure}

We start with a ``standard" inference, using the most commonly employed likelihood function (i.e., $\propto \chi^2$, with a trivial covariance matrix using only the Poisson uncertainties).  Interpolation in the model library is performed with a basic tri-linear algorithm (in this specific case, $\theta_{\ast}$ has only three dimensions).  To avoid a prominent systematic (see Section~\ref{sec:systematics}), we fix the surface gravity to $\logg = 4.29$ (with a $\delta$-function prior).  This independent prior information comes from the combination of a constraint on the mean stellar density based on exoplanet transit depth measurements and a comparison of optical photometry with stellar models in the color-magnitude diagram \citep{joshi09}.  The resulting marginal posteriors on $T_{\rm eff}$ and \Z, listed in Table~\ref{table:tests} and shown in Figure~\ref{fig:PHOENIX_posterior}, are remarkably narrow -- unbelievably so, given how subtly the spectrum changes over such small parameter deviations.          

For the second test, we increase the complexity of the covariance matrix by introducing the global kernel treatment discussed in Section~\ref{subsec:global_covariance}. We find non-negligible amplitudes and correlation lengths for these kernels, as would be expected for a typical correlated residual spectrum.  With respect to the standard inference, the uncertainty associated with $\teff$ has increased by roughly a factor of three, but the $\Z$ posterior is only marginally broadened (by $\sim$50\%).  Upon closer inspection of the latter, it becomes clear that the posterior has an artificially sharp peak located at a grid point in the model library ($\Z = -0.5$).  This `noding' is an artifact of naive interpolation over a sparsely-sampled dimension in the library grid; when the uncertainty in the interpolation itself constitutes a significant fraction of the total error budget, the fit will be driven toward grid points (where the interpolation error is naturally minimized; see also \citealt{cottaar14}).  To mitigate this behavior, we need to employ an interpolation scheme that properly incorporates this kind of uncertainty.        

\capstartfalse
\begin{deluxetable}{clc|r@{ $\pm$ }lr@{ $\pm$ }l}[!bh]
\tablecaption{\label{table:tests} Demonstration Tests for WASP-14}
\tablehead{\colhead{Test} & \colhead{Interp} & \colhead{$\mathsf{C}$} & \multicolumn{2}{c}{$\teff$} & \multicolumn{2}{c}{$\Z$} }
\startdata
(1) & linear & trivial & 6280 & 5 &  $-$0.471 & 0.004 \\
(2) & linear & + $\mathcal{K}^{\rm G}$ & 6297 & 16 &  $-$0.500 & 0.006 \\
(3) & emulator & + $\mathcal{K}^{\rm G}$ & 6281 & 26 &  $-$0.482 & 0.012 \\
(4) & emulator & + $\mathcal{K}^{\rm G} + \mathcal{K}^{\rm L}$ & 6301 & 29 &  $-$0.431 & 0.012
\enddata
\tablecomments{The best-fit parameter values (peak of the marginal posteriors) and associated (1\,$\sigma$) uncertainties (68.3\%\ confidence intervals) for the four tests of increasing complexity described in the text.  Note that $\logg$ is fixed to 4.29 \citep[cf.,][]{joshi09}.}
\end{deluxetable}
\capstarttrue

\begin{figure*}[!t]
\includegraphics{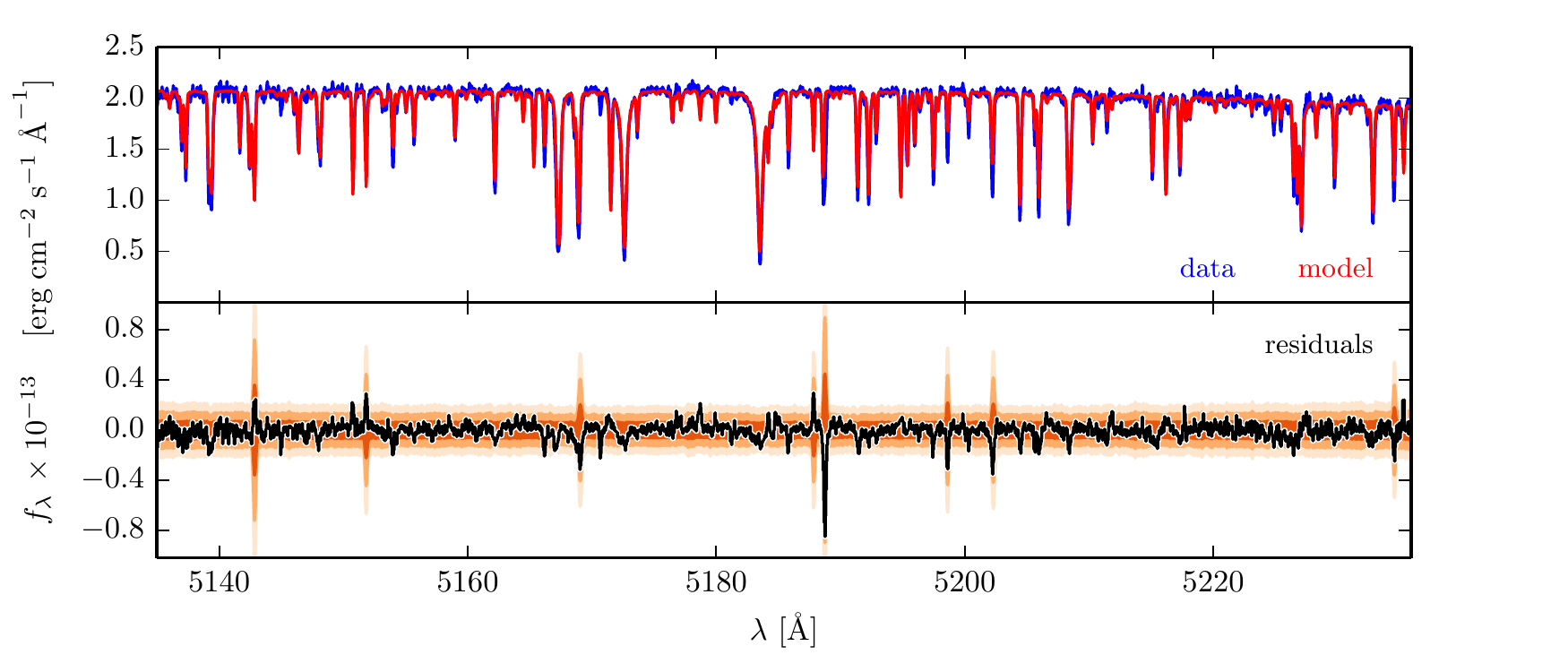}
\vspace{-0.3cm}
\figcaption{({\it top}) A representative segment of the TRES spectrum of WASP-14 ({\it blue}), overlaid with a \PHOENIX\ model ({\it red}) generated by drawing parameters from the posterior distribution (under the assumption of a fixed $\log g = 4.29$).  ({\it bottom}) The corresponding residual spectrum overlaid on contours representing the distributions of a large number of random draws from the covariance matrix (the shading is representative of the 1, 2, and 3\,$\sigma$ spreads of that distribution of draws), as in Fig.~\ref{fig:matrix}.  Note the utility of local patches of increased residual variance in accounting for outlier features, which are introduced by the local covariance kernels described in Sect.~\ref{subsec:local_covariance}.   \label{fig:PHOENIX_residuals}}
\vspace{0.2cm}
\end{figure*}

Therefore, in a third test we implement the Bayesian emulator described in Appendix~\ref{sec:Appendix} to propagate uncertainty in the interpolation.  This procedure successfully avoids the `noding' behavior in $\Z$, and inflates the associated uncertainty by a factor of 2.5 compared to the ``standard" inference approach.  The uncertainty on $\teff$ is now 5$\times$ larger than in the original test.  

Finally, in a fourth test we fold in the methodology for the local covariance kernels described in Section~\ref{subsec:local_covariance}.  This has little effect on the widths of the parameter posteriors ($\lesssim$10\%\ increase), but does shift their peaks to slightly higher values in both $\teff$ and $\Z$.  We suspect this is likely driven by a bias in the inference of $\Z$, produced because the \PHOENIX\ models tend to have more `outlier' spectral lines with {\it over-predicted} line depths.  Without the local covariance kernels to downweight these outliers, the models tend toward lower metallicity to account for them.  But when the local kernels are included, this bias is reduced and a more appropriate higher $\Z$ value is inferred.  Figure~\ref{fig:PHOENIX_residuals} demonstrates how well the modeling framework can match the character of the residual spectrum when employing the sophisticated covariance matrix (test 4) advocated here.

\subsection{Gl~51}


A moderate resolution ($R\approx2,000$) near-infrared spectrum of Gl~51 was obtained on 2000 Nov 6 using the SPEX instrument \citep{rayner03} on the 2.3\,m NASA Infrared Telescope Facility (IRTF).  SPEX is a cross-dispersed echelle spectrograph that covers the red-optical to thermal-infrared spectrum (0.7--5.5\,$\mu$m) in two settings.  These data were obtained as part of the IRTF spectral standard library project \citep{cushing05,rayner09}, and were processed through the well-vetted {\tt Spextool} reduction pipeline \citep{cushing04,vacca03} to deliver a fully calibrated spectrum.  At 2.1\,$\mu$m, the S/N is $\sim$400 per resolution element.

\capstartfalse
\begin{deluxetable}{clc|r@{ $\pm$ }lr@{ $\pm$ }l}[!bh]
\tablecaption{\label{table:tests_gl51} Demonstration Tests for Gl~51}
\tablehead{\colhead{Test} & \colhead{Interp} & \colhead{$\mathsf{C}$} & \multicolumn{2}{c}{$\teff$} & \multicolumn{2}{c}{$\Z$} }
\startdata
(1) & linear & trivial & 3256 & 3 &  0.89 & 0.01 \\
(2) & linear & + $\mathcal{K}^{\rm G}$ & 3022 & 35 &  0.00 & 0.03 \\
(3) & emulator & + $\mathcal{K}^{\rm G}$ & 3230 & 30 &  0.27 & 0.03 \\
(4) & emulator & + $\mathcal{K}^{\rm G} + \mathcal{K}^{\rm L}$ & 3180 & 35 &  0.28 & 0.04
\enddata
\tablecomments{The best-fit parameter values and associated (1\,$\sigma$) uncertainties (as in Table~\ref{table:tests}) for the four tests of increasing complexity described in the text.  Note that $\logg$ is fixed to 5.0 \citep[cf.,][]{rojas-ayala12}.}
\end{deluxetable}
\capstarttrue

\begin{figure}[!b]
  \includegraphics[width=0.5\textwidth]{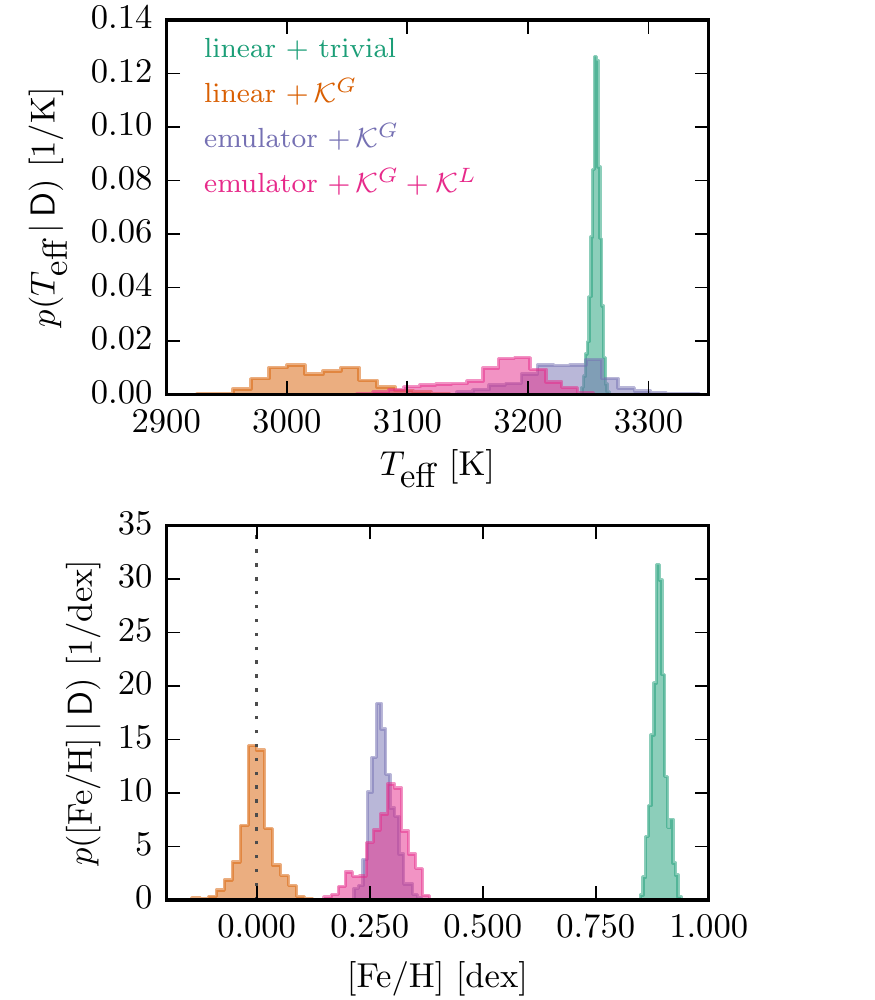}
  \vspace{-0.5cm}
  \figcaption{The marginal posterior probability distributions for the WASP-14 $\teff$ and $\Z$ based on the {\sc PHOENIX} model library, for various levels of model complexity, including: (1) a simple linear interpolation scheme and trivial covariance matrix ({\it blue-green}); (2) including global covariance terms from Gaussian process kernels ({\it orange}); (3) employing a Bayesian emulator for more appropriate interpolation ({\it purple}); and (4) also including local covariance kernels to downweight systematic outlier spectral lines ({\it magenta}). 
\label{fig:Gl51_posterior}}
\end{figure}

\begin{figure*}[!ht]
  \includegraphics{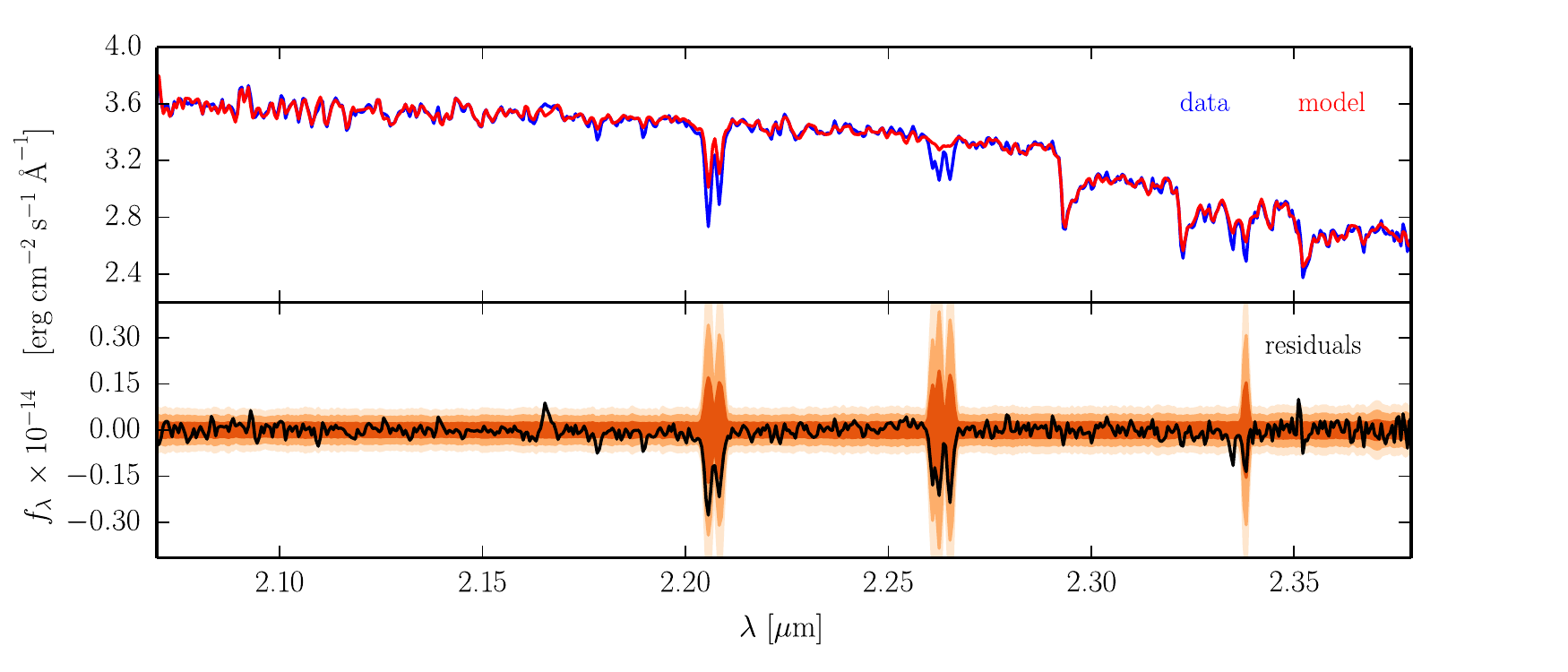}
\vspace{-0.3cm}
  \figcaption{The $K$-band SPEX spectrum of Gl~51 ({\it blue}) compared with a {\sc Phoenix} model ({\it red}) generated by drawing parameters from the inferred posterior distribution. ({\it bottom}) The residual spectrum along with contours representing the distributions of a large number of random draws from the covariance matrix (the shading is representative of the 1, 2, and 3\,$\sigma$ spreads of that distribution of draws), as in Fig.~\ref{fig:PHOENIX_residuals}.  Note how the `outlier' features (\ion{Na}{1} at 2.21\,$\mu$m and \ion{Ca}{1} at 2.26\,$\mu$m) are identified and treated by the local covariance kernels.
\label{fig:Gl51_residuals}}
\vspace{0.2cm}
\end{figure*}

Modeling late-type stellar atmosphere structures and their spectra is considerably more complex than for Sun-like stars, due to lingering uncertainties in the atmosphere physics and molecular opacities.  Especially confounding is the presence of complex condensates (clouds) at the coolest temperatures \citep{allard13}, making it considerably more challenging to determine (sub-) stellar properties \citep{rajpurohit14}.  Various approaches have been taken to infer the key parameters in the face of these difficulties, including iteratively masking regions with poor spectral agreement \citep[e.g.,][]{mann13}.  Astutely, \citeauthor{mann13}~note that such a scheme may exclude regions of the spectrum that contain intrinsically useful information for discriminating between physical properties, and that a more sophisticated approach would weight each spectral region based on its consistency with the data.  The modeling framework that we have constructed here does exactly that.

As another demonstration of this framework, we carried out the sequence of tests outlined in the previous section for the $K$-band portion of the SPEX spectrum of Gl~51.  Following the analysis of similar data for this star by \citet{rojas-ayala12}, we fix the surface gravity to $\logg = 5.0$ based on a comparison with standard stellar evolution models.  The test results are listed in Table~\ref{table:tests_gl51}; the posteriors for $\teff$ and $\Z$ are shown together in Figure~\ref{fig:Gl51_posterior}.  Like WASP-14, we find that a more appropriate treatment of the covariance matrix results in a substantial broadening of the parameter posteriors; the uncertainties on $\teff$ and $\Z$ are inflated by a factor of $\sim$10 and 4, respectively.  

However, in this case the parameter values (posterior peaks) also exhibit substantial movement along the sequence of tests.  The underlying cause of this behavior lies with the \ion{Na}{1} and \ion{Ca}{1} resonance line depths, which are systematically under-predicted in the {\sc Phoenix} library (even for high metallicities; see also \citealt{rojas-ayala12}).  \citet{rajpurohit10} suggest that these discrepancies may be the consequence of inaccurate atomic data (oscillator strengths and/or opacities).  In the first test with a trivial covariance matrix, these `outlier' lines drive the model to favor a very high $\Z$.  But when we consider the more sophisticated versions of $\vC$ that employ Gaussian processes to treat correlated residuals, the contribution of these features to the likelihood calculation is reduced, and therefore $\Z$ returns to a more appropriate range.  Because this portion of the spectrum has only these two outlier features, their influence can be mitigated either with a larger global covariance kernel amplitude ($a_{\rm G}$), or with a smaller $a_{\rm G}$ but significant contributions from local covariance kernels (which explains why there is little difference between the posteriors in the third and fourth tests in the sequence).  For reference, Figure~\ref{fig:Gl51_residuals} compares the observations with draws from the posterior distribution for the advocated modeling approach (corresponding to the fourth test).


The methodology behind the likelihood calculations we have developed could prove especially useful for spectroscopic inferences of the parameters of cool stars like Gl~51, where substantial uncertainties in their more complex atmospheres will naturally produce systematic deviations between models and data.  However, many of those discrepancies will be manifested in molecular features, which likely result in considerably more complex residual structures than noted here \citep[e.g., the TiO bands in the red-optical; see][their Fig.~9]{mann13}.  The overall framework we have employed should still function, although more appropriate local covariance kernels may need to be developed to capture the different nature of these outliers.  For example, one might employ hybrid kernels (like the product of a truncated exponential and a \matern\ kernel) or empirically-motivated parametric shapes (e.g., a saw-tooth pattern) to provide a better representation than a simple Gaussian feature.

\subsection{Synopsis and Systematics} \label{sec:systematics}

The results of the sequence of tests in the previous two sections illustrate some key issues in the spectroscopic inference of stellar parameters.  First, the residual spectra derived from (typically) imperfect models exhibit correlated structure (e.g., see Fig.~\ref{fig:class0}) that cannot be explained well with a trivial (diagonal) covariance matrix.  If that naive assumption is made (as is usually the case), the resulting posteriors are unrealistically narrow and may end up being biased (particularly for $\Z$ or for cases influenced by prominent `outlier' lines).  

This issue of implausibly small formal uncertainties has long been recognized in the stellar spectroscopy community.  The standard solution has been to add (in quadrature) a  `floor' contribution, imposed independently on each parameter and meant to be representative of the systematics (e.g., see \citealt{torres12} or \citealt{schonrich14} for clear and open discussions of this approach).  The key problems with this tactic are that these systematics are in reality degenerate (and so should not be applied independently) and that they dominate the uncertainty budget, but are in a large sense arbitrary -- they are not self-consistently derived in the likelihood framework.  Our goal here has been to treat one aspect of this systematic uncertainty budget internal to the forward-modeling process, by employing a non-trivial covariance matrix that accounts for generic issues in the pixel-by-pixel inference problem.  Given the results above, we have demonstrated that this procedure successfully accounts for a substantial fraction of the (empirically motivated) {\it ad hoc} systematic `floor' contribution typically adopted in inference studies.    

However, although a likelihood function that can properly account for the character of the residuals is important, it does not by itself treat {\it all} of the important kinds of systematics in the general spectroscopic inference problem.  In future work that can build on the flexible likelihood formalism we have advocated here, there are three other important sources of systematic uncertainty that should be considered: (1) data calibration; (2) optimized parameter sensitivity; and (3) model assumptions, or flexibility.  We discuss each of these issues briefly, with attention paid to potential remedies that fit within the likelihood framework developed here. 

Perhaps the most familiar source of systematics lies with issues in the data calibration.  In the idealized case of perfect calibration, the physical parameters inferred from different observations of the same (static) source should be indistinguishable.  But given the complexity of a detailed spectroscopic calibration, that is not typically the case in practice.  The common approach to quantify the systematic uncertainties contributed by calibration issues is to compare the inferences made using different spectra (e.g., from different instruments and/or observations).  The final parameter values are usually presented as an average of these separate inferences, with the uncertainties inflated by adding in quadrature some parameter-independent terms that account for their dispersion.  The more appropriate way of combining these inferences is to model the individual spectra simultaneously in a hierarchical framework like the one discussed in Section~\ref{sec:method}: in that way, the dispersion is appropriately propagated into the parameter uncertainties while any intrinsic degeneracies are preserved (which is not possible in the standard `weighted average' approach).  Ultimately, one could also introduce some empirically-motivated nuisance parameters that are capable of forward-modeling imperfections in the data calibration, similar to the approach adopted in Section~\ref{subsec:postprocess} (e.g., see Fig.~\ref{fig:chebyshev}).

Another important source of systematic {\it bias} comes from the fact that certain physical parameters have only a relatively subtle effect on the spectrum.  Stellar spectroscopists are familiar with this being an issue when inferring the surface gravity, $\logg$, since it is primarily manifested as low-level modifications to the wings of certain spectral lines like Mg~b and in the equivalent widths of lines from singly-ionized elements like \ion{Ti}{2} and \ion{Fe}{2}.  When modeling data with a large spectral range, the effects of varying $\logg$ are small compared to the residuals introduced by the many other model imperfections.  Consequently, the surface gravity will not be constrained well, and inferences on $\logg$ (and therefore other degenerate parameters) can be substantially biased.  As an example, when fitting the WASP-14 data in Section~\ref{subsec:wasp} without prior information on the surface gravity, we find a shift of $\sim$0.9 dex to lower $\logg$ (and accompanying shifts in $\teff$ and $\Z$).  If we instead use a customized version of the \citet{castelli04} models designed to more accurately reproduce this part of the optical spectrum for Sun-like stars (as employed by {\tt SPC}; \citealt{buchhave12}), there is still a 0.2 dex shift compared to the independent, accurate constraints from the transiting planet \citep{joshi09}.  Similar work with larger samples indicate a typical scatter in the $\logg$ values inferred solely from spectra relative to independent, accurate constraints from other data ($\sim$0.5 dex; \citealt{cottaar14,schonrich14}).        

There are two commonly utilized, and not mutually exclusive, approaches to mitigating this kind of bias.  First is the judicious use of a prior, based on either independent and accurate measurements (e.g., from asteroseismology, dynamical masses and distances, etc.) or stellar evolution models (as is demonstrated here).  Of course, such information is unfortunately not always readily available for the target of interest.  A second approach is to severely limit the spectral range of the data being modeled, focusing primarily on those spectral features especially sensitive to the parameter of interest.  But that carries its own risk, since the models derived from the inferred posteriors might well be wildly inconsistent with the rest of the spectrum.  Recently, \citet{brewer15} proposed a sophisticated, iterative approach that apparently resolves this issue, employing a sequence of conditional inferences based on sets of specific spectral features that are especially sensitive to individual physical parameters.  This seems like a promising component for future inclusion in the likelihood framework we have developed in Section~\ref{sec:method}.

Finally, and perhaps most significant, there are also sources of systematic bias and uncertainty introduced by limitations in the synthetic stellar models themselves.  Different models make varied assumptions in their treatments of the atmosphere structures, boundary conditions (e.g., convection), fundamental atomic/molecular data (e.g., opacities), and radiative transfer.  Taken together, these variations produce notably different model spectra for the same values of the physical parameters.  As a benchmark for estimating the scope of this source of bias, we re-performed the inference described in the fourth test of Section~\ref{subsec:wasp}, but using the customized \citet{castelli04} model library instead of the \citet{husser13} library.  The resulting inferences for $\teff$ and $\Z$ are in excellent agreement with those derived by \citet{torres12} using the {\tt SPC} method, but are shifted by 150\,K (higher) and 0.15\,dex (higher), respectively, compared to the {\sc Phoenix} results. While the relevant physics included in these models is very similar for these temperatures and the inferred stellar parameters are similar in an absolute sense, it is still striking that the systematic shift between models is several times larger than the statistical uncertainties derived from our likelihood function. At this point, there is little to be done to rectify these model-dependent differences; in the future, one hopes that the model inputs can be refined based on feedback from the data (see Sect.~\ref{sec:discussion}).  Any inferences of physical parameters should only be considered in the context of the assumed models.

Aside from these different assumptions and inputs, the limited {\it flexibility} of these models certainly also contributes to the systematic uncertainty budget, and is possibly also a source of systematic bias.  Model spectral libraries typically have neglected dimensions in parameter-space that, if made available, would be expected to broaden and possibly shift the posteriors for the primary physical parameters.  One typical example lies with element-specific abundance patterns, often distilled to the enhancement of $\alpha$-elements (i.e., [$\alpha$/Fe]).  If the target star has a non-zero [$\alpha$/Fe] (an enhancement or deficit relative to the solar ratios), but is fit with a single, global metallicity pattern, it is not clear that the sophisticated covariance formalism developed here would be capable of appropriately capturing such residual behavior.  Another prominent example of an important hidden parameter dimension is the microturbulence, which for some spectral types and spectral resolution may impact the spectrum in a similar way as the surface gravity \citep[and may be partly responsible for the $\logg$ bias discussed above;][]{gray01}.  To mitigate the resulting deficiencies in precision (and potentially accuracy) on the inference of other parameters, we would ideally employ libraries or modeling front-ends that can incorporate some flexibility in these hidden (i.e., ignored) dimensions of parameter-space (e.g., individual elemental or group-based abundance patterns, microturbulence, etc.).     

\section{Discussion} \label{sec:discussion}

Astronomers exploit spectroscopy to retrieve physical information about their targets.  Ideally, such inferences are made with the maximal precision afforded by the measurement noise, and accurately reflect the uncertainties with minimal systematic bias.  But in practice, the spectral models used as references are never perfect representations.  Even modest mismatches between data and model can propagate substantial systematic uncertainty into the inference problem.  In high-sensitivity applications (e.g., stellar and exoplanetary astrophysics), ignoring these systematics can give a false sense of both precision and accuracy in the inferences of key parameters.  Typically, the more egregious of these imperfections are ``mitigated" by dismissal (explicitly not considering a subset of the data; e.g., masking, clipping).  Rarely, they are confronted directly with painstaking, computationally expensive fine-tuning of more general (nuisance) parameters in the model (e.g., oscillator strengths, opacities), albeit only over a very limited spectral range and region of physical parameter-space.

We have presented an alternative approach to dealing with this fundamental issue, grounded in a generative Bayesian framework.  The method advocated here constructs a sophisticated likelihood function, employing a non-trivial covariance matrix to treat the correlated pixel-to-pixel residuals generated from intrinsically imperfect models.  That matrix is composed of a linear combination of {\it global} (stationary) and {\it local} (non-stationary) Gaussian process kernels, which parameterize an overall mild covariance structure as well as small patches of highly discrepant outlier features. In the context of a given model parameterization (i.e., synthetic spectral library, or a more complex and flexible model generator), the framework we have developed provides a better inference than the standard $\chi^2$ (or cross-correlation) comparison. We have built up a series of tests that demonstrates how the emulator, global kernels, and local kernels affect the nature of the inference on the stellar parameters. To demonstrate how the framework is used, we determined the surface parameters of main-sequence stars with mid-F and mid-M spectral types from high-S/N optical and near-infrared data, with reference to pre-computed model libraries (Sect.~\ref{sec:examples}).  The source code developed here is open and freely available for use: see \url{http://iancze.github.io/Starfish}.

The novelty of employing this kind of likelihood function in the spectroscopic inference problem is that the treatment of data--model mismatches (in essence, the fit quality) is explicitly built into the forward-modeling framework.  This offers the unique advantage that discrepant spectral features (outliers), which may contain substantial (even crucial) information about the parameters of interest, can still effectively propagate their useful information content into the posteriors with a weighting that is determined self-consistently.  From a practical standpoint, this means that a larger spectral range can be used and model imperfections can be downweighted by the usage of covariance kernels. The global covariance framework provides more appropriate estimates of the posterior probability distribution functions (i.e., the precision or uncertainty estimates) for the model parameters.  The automated identification and disciplined downweighting of problematic ``outlier" spectral lines (those that cannot be reproduced with any combination of the model parameters) with local covariance kernels can prevent them from overly influencing (and possibly biasing, especially in cases with few spectral features available) the inferences. In many cases, the underlying physical problem lies with incorrect (or inaccurate) atomic and/or opacity data used in the models.  In this sense, the posteriors of the hyperparameters of the local covariance kernels can actually indicate in what sense and scale these inputs need to be modified to better reproduce observational reality.  

The approach we describe is generally applicable to any spectroscopic inference problem (e.g., population synthesis in unresolved star clusters or galaxies, physical/chemical models of emission line spectra in star-forming regions, etc.).  Moreover, it has the flexibility to incorporate additional information (as priors) or parametric complexity (if desired), and could be deployed as a substitute for a simplistic $\chi^2$ metric in already-established tools (e.g., {\tt SME}). Another potential application might be in the estimation of radial velocities using traditional Doppler-tracking pipelines for exoplanet or binary star research.  Poorly modeled micro-tellurics can lead to incorrect measurements of radial velocities for certain contaminated chunks of the spectrum, causing them to give unrealistically precise but biased velocity measurements.  A flexible noise model would broaden the posteriors on these points and allow them to be combined into a more accurate systemic velocity. 

Ultimately, the benefits of employing covariance kernels to accommodate imperfect models could be extended well beyond modeling the spectra of individual targets.  In principle, the approach we have described here can be used to systematically discover and quantify imperfections in spectral models and eventually to build data-driven improvements of those models that are more appropriate for spectroscopic inference. By fitting many stellar spectra with the same family of models, we can catalog the covariant structure of the fit residuals -- especially the parameters of the local covariance kernels -- to collate quantitative information about where and how the models tend to deviate from observational reality.  That information can be passed to the spectral synthesis community, in some cases enabling modifications that will improve the quality of the spectral models.  On a large enough scale, this feedback between observers and modelers could be used to refine inputs like atomic and molecular data (oscillator strengths, opacities), elemental abundance patterns, and perhaps the stellar atmosphere structures. If one has access to the radiative synthesis process that generates the model spectra, there are many possible means to improve their quality.  In particular, a process of history matching can be used to rule out regions of parameter space where the models do not fit well (e.g., for a use in galaxy formation simulations, see \citet{vernon10}). For example, if one had full control over the radiative synthesis code, stellar structure code, and atomic line database, one could improve the performance of the spectral emulator by ruling out regions of parameter space for these separate components that are inconsistent with a collection of observed spectra, such as a set of standard stars spanning the full range of spectral classifications. 

In a similar vein, we could also simultaneously use several synthetic spectral libraries to infer the stellar parameters while also identifying discrepant regions of the spectrum.  A treatment using multiple synthetic libraries would likely reveal  interesting correlations between model discrepancies, such as a specific signature among many lines (e.g. deviations in spectral line shape that cannot be explained by variations in $\vt$).  Conversely, if a discrepant feature is seen for all models, it could be due to either an anomaly with the given star (e.g., a chromospheric line due to activity or perhaps an intervening interstellar absorption line) or a correlated difficulty among all models (e.g., an incorrect atomic constant). 

Alternatively, this kind of feedback could be used to make data-driven modifications to the already existing models, creating a new semi-empirical model library.  This could be accomplished by linking the parameters of the covariance kernels while fitting many stars of similar spectral type in a hierarchical Bayesian model, which would add confidence to the assessment that certain spectral features are {\it systematic} outliers and offer general quantitative guidance on how to weight them in the likelihood calculation.  Rather than simply assembling an empirical spectral library using only observations, this combined machine-learning approach would naturally provide a physical anchoring for the key physical parameters, since they are reflected in the spectra based on the physical assumptions in the original models.  This kind of large-scale analysis holds great promise in the (ongoing) era of large, homogeneous high resolution spectroscopic datasets (e.g., like those being collected in programs like the APOGEE and HERMES surveys; \citealt{nidever12, zucker12}), since they provide enormous leverage for identifying and improving the underlying model systematics. \\

\acknowledgments  The authors would like to acknowledge the following people for many extraordinarily helpful discussions and key insights: Daniel Foreman-Mackey, Guillermo Torres, David Latham, Lars Buchhave, John Johnson and the ExoLab, Daniel Eisenstein, Rebekah Dawson, Tom Loredo and the ExoStat group, Allyson Bieryla, and Maxwell Moe.  Two anonymous reviewers provided encouraging and very useful comments on the manuscript draft that greatly improved its clarity and focus.  IC is supported by the NSF Graduate Fellowship and the Smithsonian Institution. K.M. is supported at Harvard by NSF grant AST-1211196. 
This research made extensive use of Astropy \citep{astropy13} and the Julia language \citep{julia12}.

\appendix

\section{Spectral Emulator for Interpolation} \label{sec:Appendix}

\begin{figure}[!t]
    \begin{center}
        \includegraphics{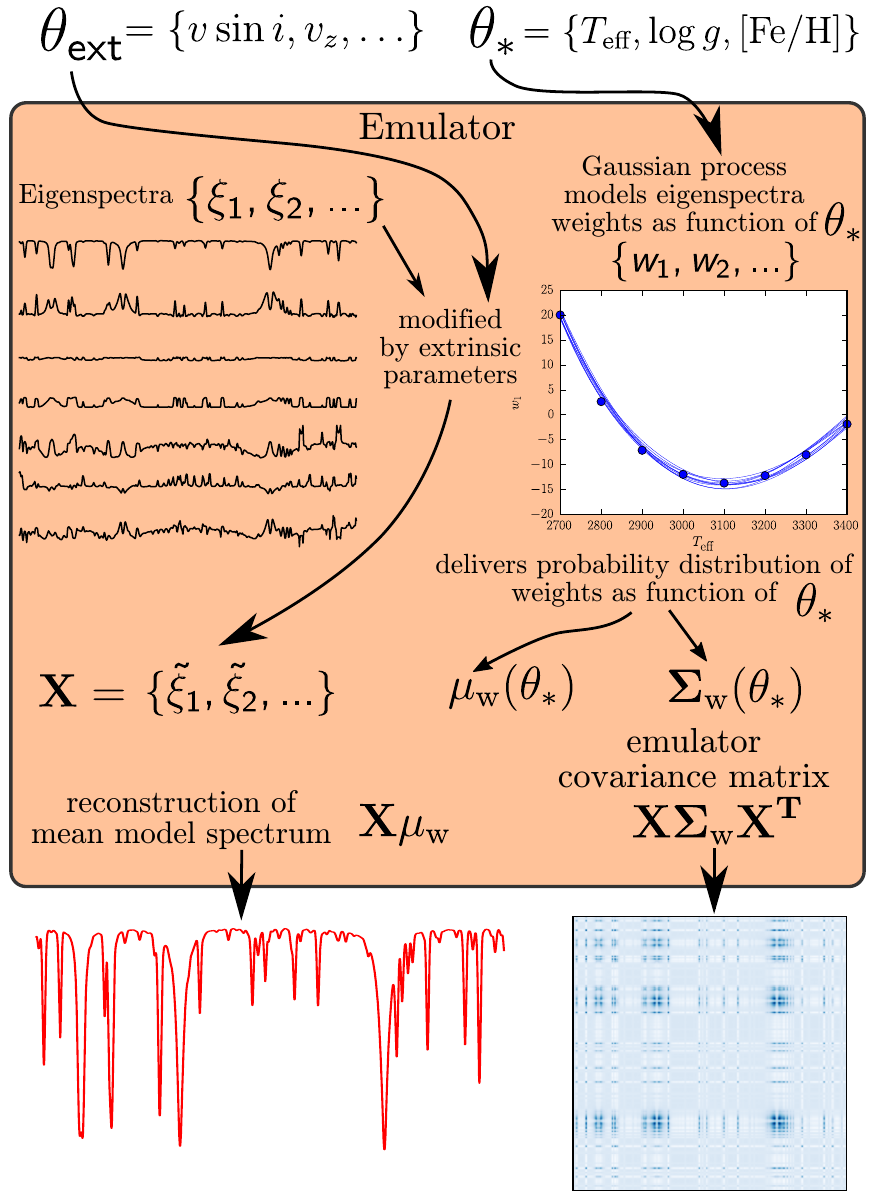}
        \figcaption{A continued flowchart explaining the contribution of the spectral emulator to the likelihood function. The synthetic library is first decomposed into an eigenspectram basis. Then, the extrinsic parameters $\vt_\textrm{ext}$ modify the eigenspectra. The intrinsic stellar parameters $\vt_\ast$ are fed into a Gaussian process (GP), which delivers a probability distribution of weights used to sum the eigenspectra. The mean weights can be used to reconstruct a mean model spectrum, while the variances of the weights are used to propagate interpolation uncertainties into the likelihood function.
        \label{fig:flowchart_appendix}}
    \end{center}
\end{figure}

The spectral emulator is designed to serve as an improved interpolator for the synthetic spectral library. Rather than a (tri-)linear interpolator, which would deliver a single spectrum for a given $\vt_\ast$, the spectral emulator delivers a probability distribution of possible interpolated spectra. In this manner, it is possible to incorporate realistic uncertainties about the interpolation process into the actual likelihood calculation. In the limit of moderate to high signal-to-noise spectra, these interpolation uncertainties can have a significant effect on the posterior distribution of $\vt_\ast$. A schematic of the emulator is shown in Figure~\ref{fig:flowchart_appendix}, which is a continuation of Figure~\ref{fig:flowchart}. Briefly, the emulator consists of a set of eigenspectra, representing the synthetic spectral library, that can be summed together with different weights to reproduce any spectrum originally in the library. To produce spectra that have $\vt_\ast$ in between $\{ \vt_\ast \}^\textrm{grid}$, the weights are modeled with a smooth Gaussian process (GP). This GP delivers a probability distribution over interpolated spectra, which can then be incorporated into the covariance matrix introduced in Section~\ref{subsec:likelihood}. Here we describe the design and construction of our spectral emulator.

\begin{figure*}[!htb]
  \begin{center}
  \includegraphics{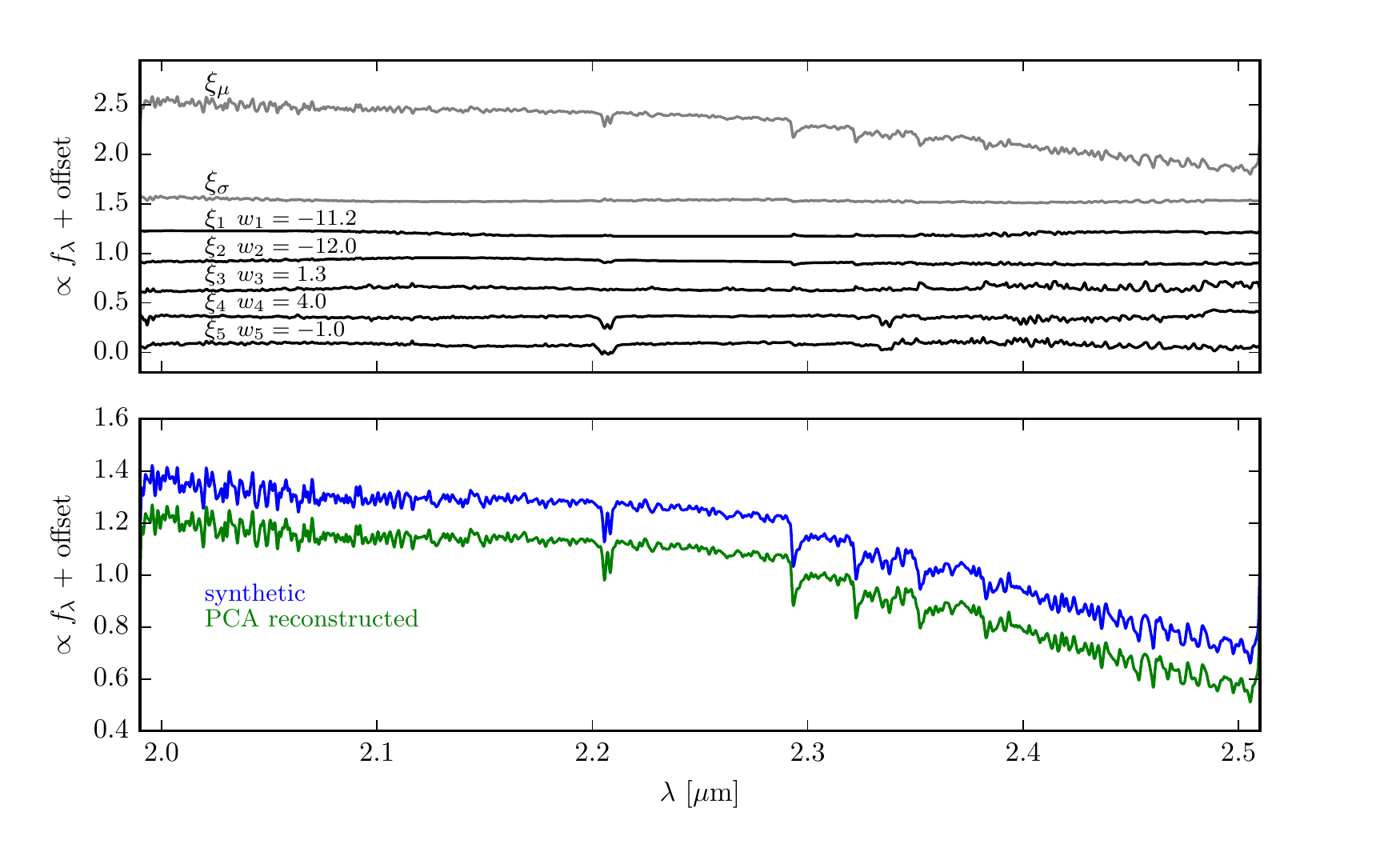}
  \vspace{-0.6cm}
  \figcaption{ (\emph{top}) The mean spectrum, standard deviation spectrum, and five eigenspectra that form the basis of the \textsc{Phoenix} synthetic library used to model Gl~51, generated using a subset of the parameter-space most relevant for M dwarfs.  (\emph{bottom}) The original synthetic spectrum from the \textsc{Phoenix} library ($\vt_\ast = [T_\textrm{eff} = 3000\,\textrm{K}, \logg = 5.0\,\textrm{dex}, \Z = 0.0\, \textrm{dex}]$) compared with a spectrum reconstructed from a linear combination of the derived eigenspectra using Eqn~\ref{eqn:reconstruct} (with the weights $w_k$ listed in the top panel figure).  \label{fig:pca_reconstruct}}
  \end{center}
\end{figure*}

Model library spectra are stored as (1-dimensional) arrays of fluxes, sampled on high resolution wavelength grids.  In the case of interest here, the sets of model parameters $\{\vt_{\ast}\}^\textrm{grid} = [\{\teff, \logg,  \Z \}]$ define the dimensions of the library grid.  The full spectral library, $f_{\lambda}(\{\vt_{\ast}\}^\textrm{grid})$, is therefore encapsulated in a 4-dimensional array.  The libraries used here have grid spacings of 0.5\,dex in $\logg$ and 0.5 dex in $\Z$; the \citet{castelli04} library steps by 250\,K in $T_{\rm eff}$, but the {\sc Phoenix} library has finer coverage in 100\,K increments.  

The first step in designing a spectral emulator is to break down the library into an appropriate basis \citep{habib07, heitmann09}.  We chose the principal component basis to decompose the library into a set of ``eigenspectra", following the techniques of \citet{ivezic13}.  Prior to this decomposition, we isolate a subset of the library (containing $M$ spectra) with parameter values that will be most relevant to the target being considered (e.g., for Gl~51, this means considering only effective temperatures below $\sim$3800\,K).  We then standardize these spectra by subtracting off their mean spectrum and then ``whiten" them by dividing off the standard deviation spectrum measured in each pixel across the grid.  The mean spectrum is 
\begin{equation}
  \xi_\mu = \frac{1}{M} \sum_{i = 1}^M f_\lambda(\{\vt_\ast \}^\textrm{grid}_i)
\end{equation}
and the standard deviation spectrum is
\begin{equation}
  \xi_\sigma = \sqrt{\frac{1}{M} \sum_{i=1}^M \bigl [ f_\lambda(\{\vt_\ast \}^\textrm{grid}_i) - \xi_\mu \bigr]^2 }, 
\end{equation}
where $\{\vt_\ast \}^\textrm{grid}$ denotes the full collection of the $M$ sets of stellar parameters under consideration in the library and $\{\vt_\ast \}^\textrm{grid}_i$ denotes a single set of those parameters drawn from this collection. Both $\xi_\mu$ and $\xi_\sigma$ are vectors with length $N_\textrm{pix}$, the same size as a raw synthetic spectrum ($f_\lambda$). In effect, all library spectra are standardized by subtracting the mean spectrum and dividing by the standard deviation spectrum
\begin{equation}
  \hat{f}_\lambda(\{\vt_\ast \}^\textrm{grid}) = \frac{f_\lambda(\{\vt_\ast \}^\textrm{grid}) - \xi_\mu}{\xi_\sigma}.
\end{equation}

The eigenspectra are computed from this standardized grid using principal component analysis \citep[PCA;][]{ivezic13}. Each eigenspectrum is a vector with length $N_\textrm{pix}$, denoted as $\xi_k$, where $k$ is the principal component index $k = \{1, 2, \ldots, m\}$. We decided to truncate our basis to the first $m$ eigenspectra, where $m$ is decided by the minimum number of eigenspectra required to reproduce any spectrum in the grid to better than 2\% accuracy for all pixels (the typical error for any given pixel is generally much smaller than this, $\lesssim 0.5\%$). As an example, the eigenspectra basis computed for Gl~51 using the \textsc{Phoenix} library is shown in the top panel of Figure~\ref{fig:pca_reconstruct}.

\begin{figure*}[!t]
\begin{center}
  \includegraphics{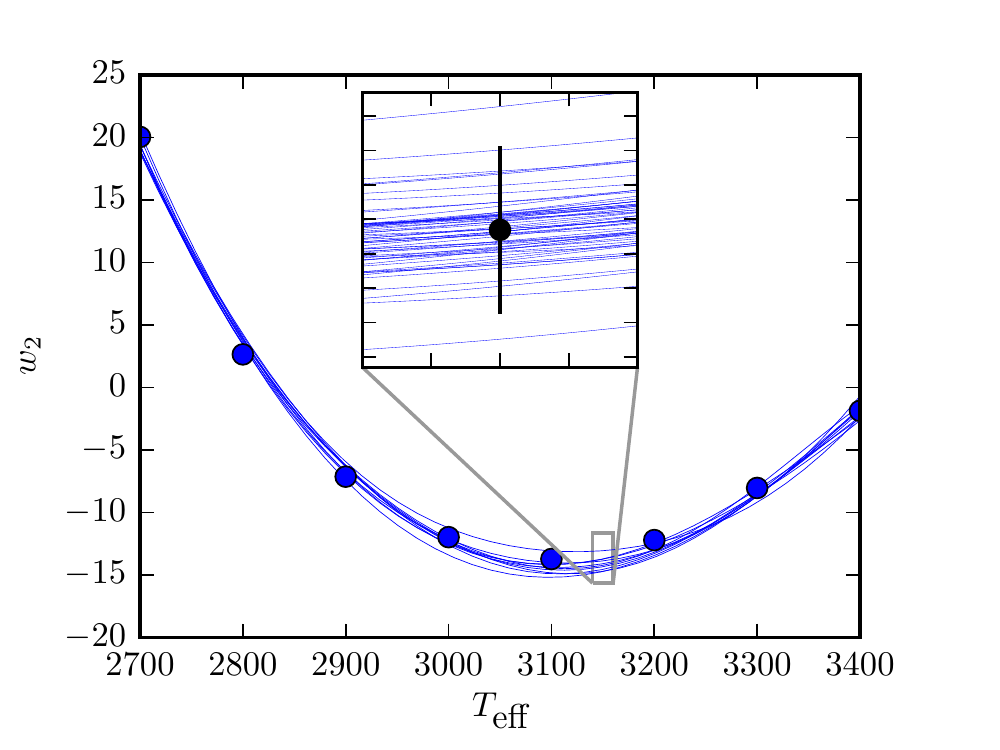}
  \includegraphics{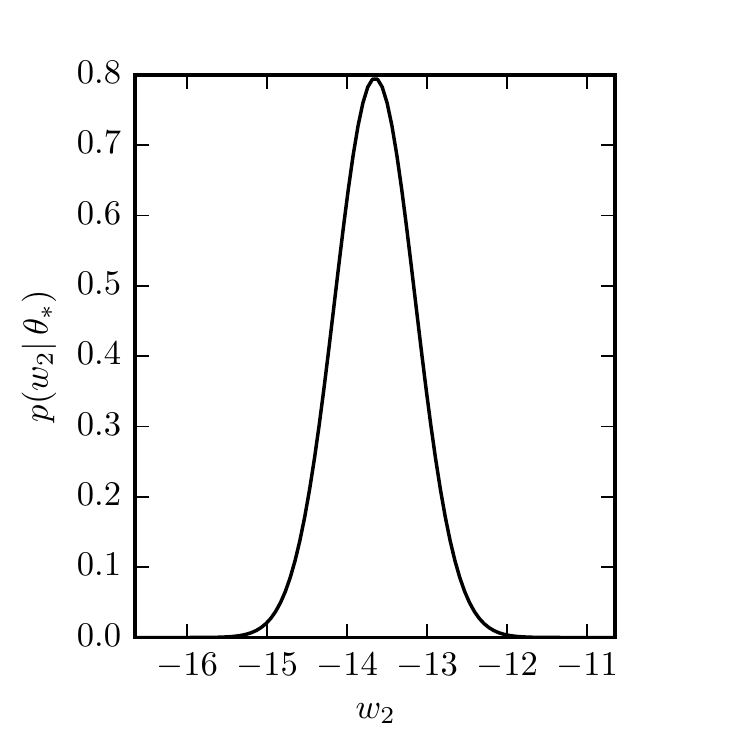}
  \figcaption{The Gaussian process modelling of the principal component weights for the Gl51 \textsc{Phoenix} spectral library. (\emph{left}) the blue dots mark the weights $\mathbf{w}_2^\textrm{grid}$ of the 2nd eigenspectrum $\xi_2$ computed at a one-dimensional slice of the spectral library for grid points with $\logg = 5.0$, $\Z = 0.0$ and various values of $T_\textrm{eff}$. In reality, the weights are a three-dimensional function of $\vt_\ast$. The thin blue lines show 50 random draws of possible functional forms described by the Gaussian process. (\emph{inset}) a zoomed portion showing the scatter in the possible functional forms. The black vertical line represents a slice through the scatter of the predicted weight value at $\vt_\ast = [\teff = 3150\,\textrm{K}, \logg = 5.0\,\textrm{dex}, \Z = 0.0\, \textrm{dex}]$. (\emph{right}) The posterior predictive probability of the collection of all weights $\mathbf{w}$ at this value of $\vt_\ast$ is completely described by Eqn~\ref{eqn:weight_conditional}, allowing us to analytically marginalize over all probable values of the weights, and thus marginalize over all probable spectral interpolations.
\label{fig:GP_interp}}
\end{center}
\end{figure*}

Using the principal component basis, we can lossily reconstruct any spectrum from the library with a linear combination of the eigenspectra
\begin{equation}
  f_\lambda(\{\vt_\ast \}^\textrm{grid}_i) \approx \xi_\mu + \xi_\sigma \sum_{k=1}^m w_k(\{\vt_\ast \}^\textrm{grid}_i) \, \xi_k
  \label{eqn:reconstruct}
\end{equation}
where $w_k$ is the weight of the $k^{\rm th}$ eigenspectrum.  These weights are 3-dimensional scalar functions that depend on the stellar parameters $\vt_\ast$.  Any given weight, which is generally a smooth function of the stellar parameters (see the left panel of Figure~\ref{fig:GP_interp}), can be determined at any grid point in the library by taking the dot product of the standardized synthetic spectrum with the eigenspectrum  
\begin{equation}
  w_k(\{\vt_\ast \}^\textrm{grid}_i) = \sum_\lambda \hat{f}_\lambda(\{\vt_\ast \}^\textrm{grid}_i) \, \xi_k.
\end{equation}
To simplify notation, we can write the collection of eigenspectra weights in a length-$m$ column 
vector
\begin{equation}
  \mathbf{w}(\vt_\ast) = 
  \begin{bmatrix}
    w_1(\vt_\ast )\\
    w_2(\vt_\ast )\\
    \vdots\\
    w_m(\vt_\ast )
  \end{bmatrix}
\end{equation}
and horizontally concatenate the eigenspectra into a matrix with $N_{\rm pix}$ rows and $m$ columns
\begin{equation}
  \mathbf{\Xi} = 
  \begin{bmatrix}
    \xi_1 & \xi_2 & \cdots & \xi_m \\
  \end{bmatrix} .
\end{equation}
Then, we can rewrite Eq.~(\ref{eqn:reconstruct}) as
\begin{equation}
  f_\lambda(\{\vt_\ast \}^\textrm{grid}_i) \approx \xi_\mu + \xi_\sigma \circ  \left ( \mathbf{\Xi} \, \mathbf{w}(\{\vt_\ast \}^\textrm{grid}_i) \right )
  \label{eqn:reconstruct2}
\end{equation}
where $\circ$ represents the element-wise multiplication of two vectors.

To recapitulate, the framework described above can be used to decompose the synthetic spectra in a model library into a principal component basis, allowing us to reconstruct any spectrum in the library as a (weighted) linear combination of $m$ eigenspectra.  The weights corresponding to each eigenspectrum are moderately-smooth scalar functions of the three stellar parameters, $\vt_{\ast}$.  Therefore, to create a spectrum corresponding to an arbitrary set of these parameters that is not represented in the spectral library, we must interpolate the weights to this new set.  In practice, it may be possible to use a traditional scheme like spline interpolation to do this directly.  However, we found that with sensitive spectra (e.g., for Gl~51 the S/N is $>$400), the uncertainty in the interpolated representation of the spectrum can constitute a significant portion of the total uncertainty budget.  This, combined with the under-sampling of the synthetic grid can cause artificial ``noding" of the posterior near grid points in the synthetic library, because the interpolated spectrum is not as good as the raw spectrum at the grid point.  Even explicitly accounting for interpolation error by doing ``drop-out" interpolation tests and empirically propagating it forward does not relieve this noding issue.  So instead, we address this problem by employing a Gaussian process to model the interpolation of the eigenspectra weights over $\vt_{\ast}$, thereby encapsulating the range of possible interpolated spectra.

Each weight is modeled by a Gaussian process for each eigenspectrum. For a single eigenspectrum with index $k$, we denote the collection of $w_k(\{\vt_\ast \}^\textrm{grid}_i)$ evaluated for all the spectra in the library as a length $M$ vector $\wg_k$.  The Gaussian process treats $\wg_k$ as a collection of random variables drawn from a joint multi-variate Gaussian distribution \citep{rasmussen05},
\begin{equation}
  \wg_k \sim \mathcal{N} \left ( \mathbf{0}, \Sg_k \right ),
\end{equation}
with $\Sg_k$ denoting the covariances. The kernel that describes the covariance matrix for this distribution is assumed to be a 3-dimensional squared exponential,
\begin{eqnarray}
  \mathcal{K}(\vt_{\ast, i}, \vt_{\ast, j} | \phi_{{\rm int},k} ) &=& a_{\rm int}^2 \exp \left [ - \frac{(T_{\textrm{eff}_i} - T_{\textrm{eff}_j})^2}{2 \, \ell_{T_\textrm{eff}}^2 } \right] \nonumber \\
  & & \times \exp \left [-\frac{(\log g_i - \log g_j)^2}{2 \, \ell_{\logg}^2} \right ] \\
  & & \times \exp \left [ -\frac{ (\Z_i - \Z_j)^2}{2 \, \ell_{\Z}^2} \right ] \nonumber,
  \label{eqn:emulator_kernel}
\end{eqnarray}
with hyperparameters $\phi_{{\rm int}, k} = \{a_{\rm int}$, $\ell_{T_{\rm eff}}$, $\ell_{\log g}$, $\ell_{\Z}$\} representing an amplitude and length scale for each dimension of $\vt_{\ast}$.  Unlike the \matern\ kernel used in Section~\ref{sec:method} (which produces a more structured behavior reminiscent of the spectral residuals), this squared exponential kernel has a smooth functional form that is more appropriate to represent the behavior of the eigenspectra weights across the library grid, as demonstrated in Figure~\ref{fig:GP_interp}.  The $M\times M$-dimensional covariance matrix is 
\begin{equation}
\mathbf{\Sigma}_k^\textrm{grid} = \mathcal{K}( \{\vt_\ast \}^\textrm{grid}, \{\vt_\ast \}^\textrm{grid} | \, \phi_{{\rm int},k}),
\end{equation}
the evaluation of the covariance kernel for all pairings of stellar parameters at library gridpoints.

Once the Gaussian processes for each $k$ are specified, we can construct the joint distribution. 
\begin{equation}
    \begin{bmatrix}
    \wg_1 \\
    \vdots \\
    \wg_m
    \end{bmatrix}
    \sim
    {\cal N}
    \left (
    \mathbf{0},
    \begin{bmatrix}
    \Sg_1 & 0 & 0 \\
    0 & \ddots & 0 \\
    0 & 0 & \Sg_m
    \end{bmatrix}
    \right )
\end{equation}
We use $\wg$ to denote the concatenation of $\wg_k$ vectors into a single length $Mm$ vector, and $\Sg$ as the $Mm \times Mm$ covariance matrix, 
\begin{equation}
\wg \sim {\cal N}(0, \Sg )
\label{eqn:weight_prior}
\end{equation}

Although we could optimize the hyperparameters of each Gaussian process independently based upon how well it reproduces the collection of weights for that eigenspectrum, ideally we would like to optimize the hyperparameters according to a metric that describes how well the emulator actually reproduces the original library of synthetic spectra.

Following \citet{habib07}, we write down a likelihood function describing how well the reconstructed spectra match the entirety of the original synthetic grid
\begin{multline}
{\cal L} ({\cal F} | \wg, \lambda_\xi) \propto \\
\lambda_\xi^{M N_\textrm{pix} /2} \exp \left [-\frac{\lambda_\xi}{2} \left ({\cal F} - \Phi \wg \right)^\trans \left ({\cal F} - \Phi \wg \right) \right]
\label{eqn:em_data_likelihood}
\end{multline}
Here, ${\cal F}$ represents a length $MN_\textrm{pix}$ vector that is the collection of all of the synthetic flux vectors concatenated end to end. The precision of the eigenspectra basis representation, or the statistical error in the ability of the emulator to reproduce the known eigenspectra is represented by $\lambda_\xi$. Because we have truncated the eigenspectra basis to only $m$ components, where $m < M$ is much smaller than the number of raw spectra in the library, the emulator will not be able to reproduce the synthetic spectra perfectly. By including this ``nugget" term in the emulator, we are also forward propagating the interpolation uncertainty for $\vt_\ast$ near or at values of $\{ \vt_\ast \}^\textrm{grid}$. We specify a broad $\Gamma$ function prior on $\lambda_\xi$ because we expect it to be well constrained by the data.
\begin{equation}
p(\lambda_\xi) = \Gamma (a_{\lambda_\xi}, b_{\lambda_\xi})
\label{eqn:gamma_priors}
\end{equation}
where shape $a_{\lambda_\xi} = 1$ and rate $b_{\lambda_\xi} = 0.0001$. To facilitate the manipulation of Eqn~\ref{eqn:em_data_likelihood}, we create a large $M N_\textrm{pix} \times Mm$ matrix that contains the all of the eigenspectra 
\begin{equation}
\vP = [I_M \otimes \xi_1, \ldots, I_M \otimes \xi_m ]
\end{equation}
where $\otimes$ is the Kronecker product. This operation creates a matrix, which, when multiplied by the vector $\wg$, enables (lossy) reconstruction of the entire synthetic library
\begin{equation}
{\cal F} \approx \vP \wg
\end{equation}
up to truncation error in the eigenspectrum basis ($\lambda_\xi$). For a given $\lambda_\xi$, the maximum likelihood estimate for Eqn~\ref{eqn:em_data_likelihood} is $\wgh = \left ( \Phi^\trans \Phi \right)^{-1} \Phi^\trans {\cal F}$. Using $\wgh$, we can factorize Eqn~\ref{eqn:em_data_likelihood} into 
\begin{multline}
{\cal L} ({\cal F} | \wg, \lambda_\xi) \propto \\
\lambda_\xi^{M m /2} \exp \left [-\frac{\lambda_\xi}{2} \left (\wg - \wgh \right)^\trans \left ( \Phi^\trans \Phi \right) \left (\wg - \wgh \right)  \right]\\
\times \lambda_\xi^{M (N_\textrm{pix} - m) /2} \exp \left [ -\frac{\lambda_\xi}{2} {\cal F}^\trans \left (I - \Phi(\Phi^\trans \Phi)^{-1} \Phi^T \right) {\cal F} \right ] \\
\end{multline}
Now, only the middle line of this distribution depends on $\wgh$, so we can reformulate this equation into a dimensionality reduced likelihood function and absorb the other terms into a modified prior on $\lambda_\xi$.
\begin{multline}
{\cal L} (\wgh \,|\, \wg , \lambda_\xi ) \propto \\
\lambda_\xi^{M m /2} \exp \left [- \frac{\lambda_\xi}{2} \left(\wg - \wgh \right)^\trans \left ( \Phi^\trans \Phi \right) \left (\wg - \wgh \right) \right]
\end{multline}
To summarize, we have reduced the dimensionality of the distribution from
\begin{equation}
{\cal L} \left ( {\cal F} \,|\, \wg, \lambda_\xi \right) = {\cal N} \left ( {\cal F} \,|\, \Phi \wg, \lambda_\xi^{-1} I_{M N_\textrm{pix}} \right)
\end{equation}
to 
\begin{equation}
{\cal L} \left (\wgh |\; \wg, \lambda_\xi \right ) = {\cal N} \left (\wg \,|\, \wgh, (\lambda_\xi \Phi^\trans \Phi )^{-1}  \right )
\label{eqn:dimensionality_reduced}
\end{equation}
Although we introduced the likelihood function in Eqn~\ref{eqn:em_data_likelihood}, we have yet to include the Gaussian processes or the dependence on the emulator parameters $\phi_\textrm{int}$. We do this by multiplying Eqn~\ref{eqn:dimensionality_reduced} with our prior distribution on the weights (Eqn~\ref{eqn:weight_prior}), 
\begin{multline}
p \left (\wg |\; \wgh, \lambda_\xi, \phi_\textrm{int} \right ) = \\
{\cal N} \left (\wg \,|\, \wgh , (\lambda_\xi \Phi^\trans \Phi )^{-1}  \right ) {\cal N}(\wg \,|\, 0, \Sg )
\end{multline}
and integrate out the dependence on $\wg$. We perform this integral using Eqn~A.7 of \citet{rasmussen05} for the product of two Gaussians, which yields
\begin{multline}
p(\wgh | \lambda_\xi, \phi_\textrm{int}) = (2 \pi)^{-M m /2} \left |(\lambda_\xi \Phi^\trans \Phi)^{-1} + \mathbf{\Sigma}_w \right|^{-1/2} \\
\times \exp \left [ -\frac{1}{2} \hat{\mathbf{w}}_d^\trans \left ((\lambda_\xi \Phi^\trans \Phi)^{-1} + \mathbf{\Sigma}_w \right)^{-1}  \hat{\mathbf{w}}_d \right] 
\end{multline}
The dimensionality reduction operation changes the priors on $\lambda_\xi$ (Eqn~\ref{eqn:gamma_priors}) to
\begin{equation}
a_{\lambda_\xi}^\prime = a_{\lambda_\xi} + \frac{M (N_\textrm{pix} - m)}{2}
\end{equation}
\begin{equation}
b_{\lambda_\xi}^\prime = b_{\lambda_\xi} + \frac{1}{2} {\cal F}^\trans \left (I - \Phi \left (\Phi \Phi^\trans \right)^{-1} \Phi^\trans \right )  {\cal F}
\end{equation}
To complete the posterior distribution for the emulator, we specify $\Gamma$ function priors on the Gaussian process length scale kernel parameters $\phi_\textrm{int}$. Typically, these priors are broad and peak at lengths corresponding to a few times the spacing between grid points, which helps the Gaussian process converge to the desired emulation behavior. The full posterior distribution is given by
\begin{equation}
p(\lambda_\xi, \phi_\textrm{int} |\, \wgh ) \propto p(\wgh |\,\lambda_\xi, \phi_\textrm{int}) \, p(\lambda_\xi, \phi_\textrm{int})
\end{equation}
where the prior is given by 
\begin{multline}
p(\lambda_\xi, \phi_\textrm{int}) = \Gamma(a_{\lambda_\xi}^\prime, b_{\lambda_\xi}^\prime) \Gamma^m(a_{\teff}, b_{\teff}) \times \\ \Gamma^m(a_{\logg}, b_{\logg}) \Gamma^m(a_{\Z}, b_{\Z}).
\end{multline}

Now that we have fully specified a posterior probability distribution, we can sample it and find the joint posteriors for the parameters $\lambda_\xi$ and the $\phi_\textrm{int}$ for all $k$ simultaneously. Once we have identified the best-fit parameters for the emulator, we fix these parameters for the remainder of the spectral fitting. 

Now, the emulator is fully specified and can be used to predict the values of the weights at any arbitrary set of stellar parameters $\vt_\ast$ by considering them drawn from the joint distribution 
\begin{equation}
    \begin{bmatrix}
        \wgh\\
        \mathbf{w}
    \end{bmatrix}
    \sim
    {\cal N} \left ( \begin{bmatrix}
                        \mathbf{0}\\
                        \mathbf{0}\\
                        \end{bmatrix} 
                        ,
                    \begin{bmatrix}
                    (\lambda_\xi \Phi^\trans \Phi) & 0\\
                    0 & 0\\
                    \end{bmatrix} + 
                    \mathbf{\Sigma}_{\wg, \mathbf{w}}
                        \right )
\end{equation}
where $\mathbf{\Sigma}_{\wg, \mathbf{w}}$ is an augmented covariance matrix that includes the point $\vt_\ast$. To simplify notation, we let
\begin{equation}
\begin{bmatrix}
V_{11} & V_{12} \\
V_{21} & V_{22}
\end{bmatrix} =
\left [
\begin{bmatrix}
    (\lambda_\xi \Phi^\trans \Phi) & 0\\
    0 & 0\\
\end{bmatrix} + 
\mathbf{\Sigma}_{\wg, \mathbf{w}} \right ]
\end{equation}
With this notation, the $M m \times M m$ matrix $V_{11}$ is the region of the covariance matrix that describes the relations between the set of parameters in the grid, $\{\vt_{\ast}\}^\textrm{grid}$. The $M m \times m$ matrix $V_{12}$ (and its transpose $V_{21}$) describe the relations between the set of parameters in the grid and the newly chosen parameters to interpolate at $\vt_\ast$. The structure of $V_{12}$ is set by evaluating ${\cal K}_i$ (Eqn~\ref{eqn:emulator_kernel}) across a series of rows of $\{\vt_{\ast}\}^\textrm{grid}$ like in $\Sg$, for $i = 1, 2, \ldots m$, and across $m$ columns of $\vt_\ast$. $V_{22}$ is a $m \times m$ diagonal matrix that represents ${\cal K}_i$ evaluated at the zero-spacing parameter pair ($\vt_\ast, \vt_\ast$), $i = 1, 2, \ldots m$. Then, to predict a vector of weights at the new location, we use the conditional probability
\begin{equation}
 p(\mathbf{w} |\, \wgh, \vt_\ast) = \mathcal{N} \left ( \mathbf{w} \left | \right .
  \mathbf{\mu}_\mathbf{w}(\vt_\ast), \mathbf{\Sigma}_\mathbf{w}(\vt_\ast) \right )
  \label{eqn:weight_conditional}
\end{equation}
where 
\begin{equation}
\mathbf{\mu}_\mathbf{w}(\vt_\ast) = V_{21} V_{11}^{-1} \wgh
\end{equation}
\begin{equation}
\mathbf{\Sigma}_\mathbf{w}(\vt_\ast) = V_{22} - V_{21} V_{11}^{-1} V_{12}
\end{equation}
These equations are also commonly referred to as kriging equations \citep{cressie93}. Though the notation is complex, the interpretation is straightforward: the probability distribution of a set of eigenspectra weights $\mathbf{w}$ is a $m$-dimensional Gaussian distribution whose mean and covariance are a function of $\vt_\ast$, conditional upon the (fixed) values of $\wgh$ and the squared exponential hyperparameters (an example for a single $w_k$ is shown in Figure~\ref{fig:GP_interp}, right panel).

If we desired actual values of the interpolated weights, for example to reconstruct a model spectrum, we could simply draw a Gaussian random variable  $\mathbf{w}$ from the probability distribution in Eq.~(\ref{eqn:weight_conditional}).  However, because we now know the probability distribution of the weight as a function of $\vt_\ast$, we can rewrite our data likelihood function (Eq.~\ref{eqn:lnlikelihood}) in such a way that it is possible to analytically marginalize over all possible values of $\mathbf{w}$, and thus all probable spectral interpolations. 

Up until this point, we have described the reconstruction of a spectrum as a linear combination of the eigenspectra that characterize the synthetic library (Figure~\ref{fig:pca_reconstruct}).  But in practice, that reconstructed spectrum must be further post-processed as detailed in Section~\ref{subsec:postprocess}.  Fortunately, because convolution is a linear operation, we can first post-process the raw eigenspectra according to $\vt_{\rm ext}$, and then represent the reconstructed spectrum as a linear combination of these modified eigenspectra without loss of information.  Unfortunately, the Doppler shift and resampling operations are not linear operations, and there will be some loss of information when trying to approximate them in this manner.  However, we find that in practice when the synthetic spectra are oversampled relative to the instrument resolution by a reasonable factor, the flux error due to resampling is smaller than 0.2\%\ across all pixels, and thus any effect of that information loss is negligible.  For notational compactness, we let $\widetilde{\xi}_\mu$, $\widetilde{\xi}_\sigma$, and $\widetilde{\mathbf{\Xi}}$ represent the post-processed eigenspectra, with an implied dependence on the current values of the extrinsic observational parameters ($\vt_\textrm{ext}$) and the polynomial nuisance parameters $(\phi_\mathsf{P})$.  Now, the model spectrum is a function of the vector of eigenspectra weights  
\begin{equation}
  \mathsf{M}(\mathbf{w}) = \widetilde{\xi}_\mu + \mathbf{X} \mathbf{w}
\end{equation}
where 
\begin{equation}
  \mathbf{X} = \widetilde{\xi}_\sigma I_{N_{\rm pix}} \widetilde{\mathbf{\Xi}}.
\end{equation}
Because the Gaussian process describes a probability distribution of the weights, we now have a distribution of possible (interpolated) models and the likelihood function (Eq.~\ref{eqn:likelihood}) is specified conditionally on the weights,
\begin{equation}
  p \bigl( \mathsf{D} |\, \mathsf{M}(\mathbf{w}) \bigr) = p( \mathsf{D} |\, \mathbf{w}) = 
  \mathcal{N} \left ( \mathsf{D} |\, \widetilde{\xi}_\mu + \mathbf{X} \mathbf{w} , \mathsf{C} \right ).
  \label{eqn:likelihood_conditional}
\end{equation}

The final task of designing the spectral emulator is to combine this data likelihood function with the posterior predictive distribution of the eigenspectra weights (Eq.~\ref{eqn:weight_conditional}) and then marginalize over the weights \begin{equation}
  p(\mathsf{D} | \vt_\ast) = \int p(\mathsf{D} | \mathbf{w}) p( \mathbf{w} | \vt_\ast) d \mathbf{w}
  \label{eqn:marginal}
\end{equation}
such that we are left with a modified posterior distribution of the data that incorporates the range of probable interpolation values for the model.  To perform this multidimensional integral, we use a convenient lemma found in \citet[their Appendix A]{gelman13}: if the probability distributions of $\mathbf{w}$ and $\mathsf{D} | \mathbf{w}$ are specified conditionally as in Eq.~\ref{eqn:weight_conditional} and \ref{eqn:likelihood_conditional}, respectively, then the marginal distribution (Eq.~\ref{eqn:marginal}) is
\begin{equation}
  p(\mathsf{D} | \vt_\ast, \vt_\textrm{obs}, \mathbf{\Phi}) = \mathcal{N} \left ( \mathsf{D} \bigl | \bigr .\, \widetilde{\xi}_\mu + \mathbf{X} \mathbf{\mu}_\mathbf{w}, \mathbf{X} \mathbf{\Sigma}_\mathbf{w} \mathbf{X}^T + \mathsf{C} \right),
\end{equation}
where the dependence on the model parameters is now made explicit.  We can couch this modified likelihood function in the form of Eqn~\ref{eqn:lnlikelihood} by rewriting
\begin{equation}
  \mathsf{M}^\prime = \widetilde{\xi}_\mu + \mathbf{X} \mathbf{\mu}_\mathbf{w}
\end{equation}
\begin{equation}
  \mathsf{R}^\prime = \mathsf{D} - \mathsf{M}^\prime 
\end{equation}
\begin{equation}
  \mathsf{C}^\prime = \mathbf{X} \mathbf{\Sigma}_\mathbf{w} \mathbf{X}^T + \mathsf{C}
  \label{eqn:modC}
\end{equation}
where $\mathsf{M}^\prime$ can be thought of as the ``mean model spectrum" given the model parameters, and the covariance matrix has been modified to account for the various probable manifestations of the model spectrum about that mean spectrum.

\bibliographystyle{yahapj}

\end{document}